\documentclass[runningheads]{llncs}

\usepackage{graphicx}
\usepackage[font=small]{caption}
\usepackage[font=scriptsize]{subcaption}
\usepackage{bbm}
\usepackage{pifont}
\usepackage{booktabs}
\usepackage{multirow}
\usepackage{graphicx}
\usepackage{hhline}
\usepackage{amsfonts}
\usepackage{amsmath}

\newcommand*\rot{\rotatebox{90}}
\newcommand{\cmark}{\ding{51}}%
\newcommand{\xmark}{\ding{55}}%
\makeatletter
\newcommand*{\centerfloat}{%
\parindent \z@
\leftskip \z@ \@plus 1fil \@minus \textwidth
\rightskip\leftskip
\parfillskip \z@skip}
\makeatother

\begin{document}
\title{Understanding Object Detection\\Through An Adversarial Lens}
\titlerunning{Understanding Object Detection Through An Adversarial Lens}

\author{Ka-Ho Chow \and Ling Liu\and\\Mehmet Emre Gursoy \and Stacey Truex \and Wenqi Wei \and Yanzhao Wu}
\authorrunning{K.-H. Chow et al.}
\institute{Georgia Institute of Technology, Atlanta, GA, USA\\
\email{\{khchow, ling.liu\}@gatech.edu,\\\{memregursoy, staceytruex, wenqiwei, yanzhaowu\}@gatech.edu}
}

\maketitle              
\vspace{-11pt}\begin{abstract}
Deep neural networks based object detection models have revolutionized computer vision and fueled the development of a wide range of visual recognition applications. However, recent studies have revealed that deep object detectors can be compromised under adversarial attacks, causing a victim detector to detect no object, fake objects, or mislabeled objects. With object detection being used pervasively in many security-critical applications, such as autonomous vehicles and smart cities, we argue that a holistic approach for an in-depth understanding of adversarial attacks and vulnerabilities of deep object detection systems is of utmost importance for the research community to develop robust defense mechanisms. This paper presents a framework for analyzing and evaluating vulnerabilities of the state-of-the-art object detectors under an adversarial lens, aiming to analyze and demystify the attack strategies, adverse effects, and costs, as well as the cross-model and cross-resolution transferability of attacks. Using a set of quantitative metrics, extensive experiments are performed on six representative deep object detectors from three popular families (YOLOv3, SSD, and Faster R-CNN) with two benchmark datasets (PASCAL VOC and MS COCO). We demonstrate that the proposed framework can serve as a methodical benchmark for analyzing adversarial behaviors and risks in real-time object detection systems. We conjecture that this framework can also serve as a tool to assess the security risks and the adversarial robustness of deep object detectors to be deployed in real-world applications.

\vspace{-5pt}
\end{abstract}

\vspace{-20pt}\section{Introduction}\vspace{-4pt}
\label{sec:introduction}
Empowered by deep structures, nonlinear activation, and high-performance GPUs, deep neural networks (DNNs) have monopolized object detection systems~\cite{redmon2018yolov3,liu2016ssd,ren2015faster}, enabling the development of many security-critical applications, such as traffic sign detection on autonomous vehicles~\cite{simon2019complexer} and intrusion detection on surveillance systems~\cite{gajjar2017human}. While deep object detection algorithms offer real-time performance with high accuracy over traditional techniques~\cite{papageorgiou1998general,viola2001rapid}, recent studies have revealed that well trained deep object detectors are vulnerable to adversarial inputs that are maliciously modified but visually imperceptible from the original benign input~\cite{chow2020tog,wei2018transferable,li2018robust,xie2017adversarial}. Table~\ref{tab:effect-examples} illustrates such vulnerabilities. With no attack, the object detector can accurately identify the person, the car, and the stop sign on the two benign images ($1$st column). However, the \emph{same} detector is fooled blindly by the adversarial examples ($2$nd-$5$th columns) that are perturbed malignantly but indistinguishable from the benign images by human-perception.
\begin{table}[t]\renewcommand{\arraystretch}{0.7}
\centering\scriptsize
\begin{tabular}{ccccc}
	\toprule
	\multirow{2}{*}[-4pt]{\textbf{\begin{tabular}[c]{@{}c@{}}{Benign}\\(No Attack)\end{tabular}}} & \multicolumn{4}{c}{\textbf{Adversarial Attacks with Different Types of Attack Specificity}} \\ \cmidrule(l){2-5}  
	& \scriptsize{\begin{tabular}[c]{@{}c@{}}{Untargeted}\\Random\end{tabular}}  & \scriptsize{Object-vanishing} & \scriptsize{{Object-fabrication}} & \scriptsize{\begin{tabular}[c]{@{}c@{}}{Object-mislabeling}\\stopsign$\rightarrow$umbrella\end{tabular}} \\\midrule 
	\includegraphics[scale=0.153]{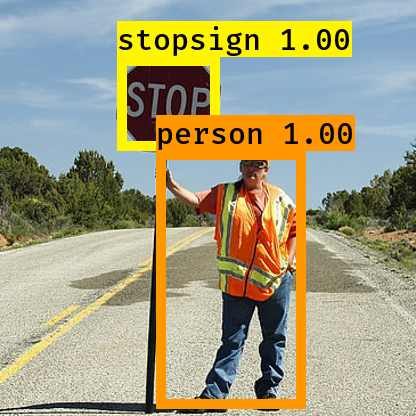} & \includegraphics[scale=0.153]{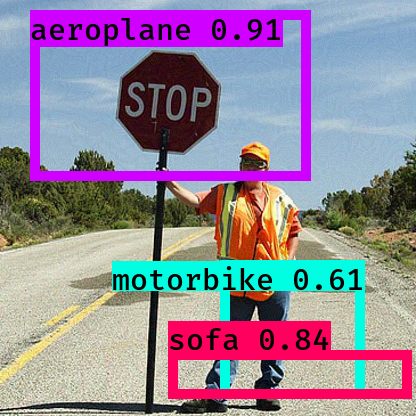} & \includegraphics[scale=0.153]{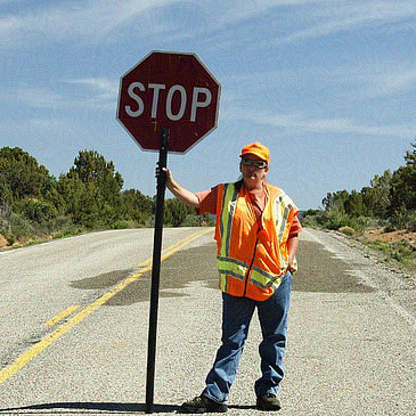} & \includegraphics[scale=0.153]{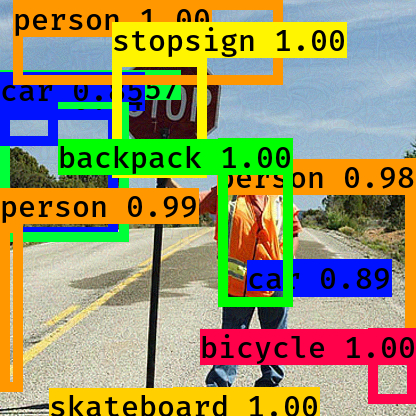} & \includegraphics[scale=0.153]{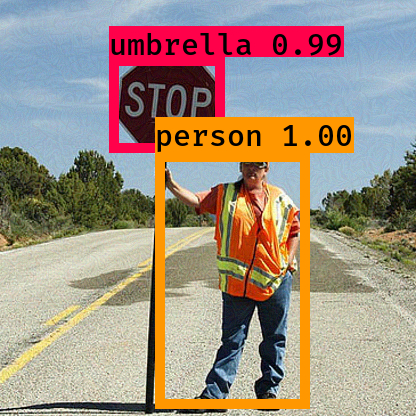} \\
	\includegraphics[scale=0.153]{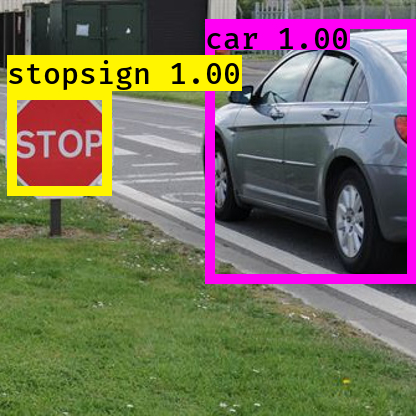} & \includegraphics[scale=0.153]{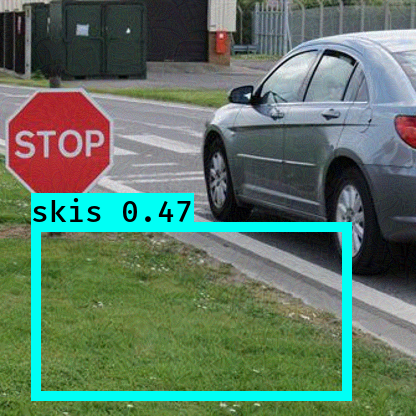} & \includegraphics[scale=0.153]{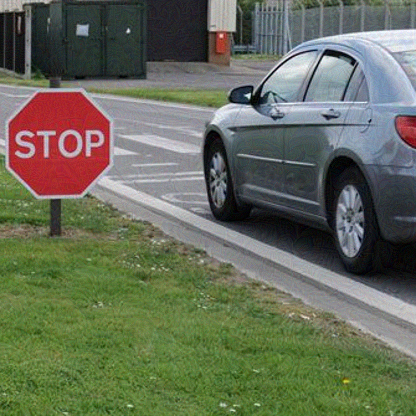} & \includegraphics[scale=0.153]{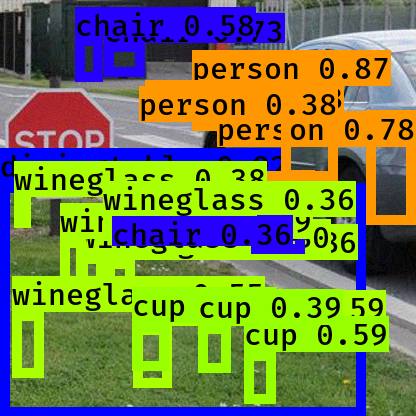} & \includegraphics[scale=0.153]{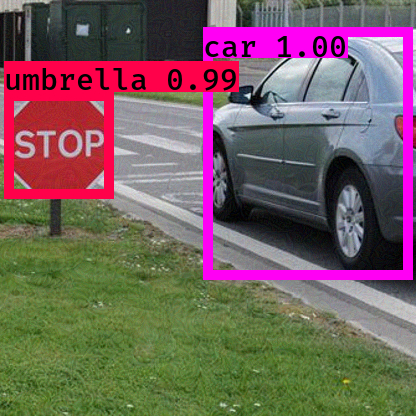} \\ \bottomrule
\end{tabular}
\vspace{4pt}
\caption{Detection on two examples by the TOG family of attacks~\cite{chow2020tog}.}
\label{tab:effect-examples}
\vspace{-24pt}
\end{table}

\vspace{-12pt}\textbf{1.1 Related Work and Problem Statement. \/} Object detection is the core task in computer vision, which takes an input image or video frame and detects multiple instances of semantic objects according to known categories~\cite{papageorgiou1998general,viola2001rapid}. Although some may view object detection as a generalization of the image classification task, a deep object detector is a multi-task learner and performs two unique learning tasks that make attacking object detection more complex and challenging than image classification: (1) Object detection should detect and identify instances of multiple semantic objects encapsulated in a single image or video frame, whereas a vanilla image classifier deals with the classification of each image into one of the known classes. (2) Object detection performs localization and classification of multiple instances of multiple semantic objects in a single image, and the localization accuracy of each instance may influence the classification accuracy of the instance. Thus, the adversarial attack techniques for image classifiers~\cite{goodfellow2014explaining,szegedy2013intriguing} are not applicable to attacking deep object detectors. The adversarial examples to attack object detection models are generated using more complex attack techniques, which compute and inject adversarial perturbations to the benign input by maximizing objectness loss, localization loss and classification loss simultaneously and iteratively~\cite{chow2020tog,wei2018transferable}. 

Existing object detection models are broadly classified into two categories: (1) the proposal-based two-phase learning and (2) the regression-based single-phase learning. The proposal-based approach uses a two-phase procedure by first detecting proposal regions with a region proposal network (RPN), and then refining them with bounding box and class label prediction. This category is dominated by Faster R-CNN~\cite{ren2015faster}, and also includes R-CNN~\cite{girshick2015fast,girshick2014rich} and Mask R-CNN~\cite{he2017mask}. The regression-based single-phase learning formulates the detection task as a regression problem. It jointly estimates the bounding box and class label of objects by directly predicting the coordinates of bounding boxes. This category is represented by YOLO~\cite{redmon2016you,redmon2017yolo9000,redmon2018yolov3,simon2019complexer} and SSD~\cite{liu2016ssd}. Moreover, different object detectors, even from the same family (e.g., Faster R-CNN), may use different neural networks as the backbone, and some additionally utilize different input resolutions~\cite{redmon2018yolov3,ren2015faster} to optimize their detection performance. Several white-box attacks are developed to attack Faster R-CNN by utilizing proposal regions, such as DAG~\cite{xie2017adversarial}, UEA~\cite{wei2018transferable}, and other similar methods~\cite{chen2018optimizing,li2018robust}. For example, DAG first assigns an adversarial label (at random) to each proposal region detected and then performs iterative gradient backpropagation to misclassify the proposals. However, DAG attack with Faster R-CNN as the victim detector cannot be applied or extended to attacking single-phase detectors, which do not use proposal regions. Similar to the black-box transfer attacks to image classifiers~\cite{papernot2016transferability}, UEA~\cite{wei2018transferable} studied the transferability of attacks by using the adversarial examples generated from a Faster R-CNN detector to attack SSD detectors.  

\textbf{1.2 Scope and Contribution.} In this paper, we develop an attack evaluation framework to rigorously analyze the vulnerabilities and security risks of deep object detection systems. The paper makes three original contributions. (1) We take a holistic approach to analyzing and characterizing adversarial attacks to object detection models from three dominant families: YOLOv3~\cite{redmon2018yolov3}, SSD~\cite{liu2016ssd}, and Faster R-CNN~\cite{ren2015faster,girshick2014rich}, including attack generalization, untargeted random attacks, targeted specificity attacks, such as object-vanishing, object-fabrication, and targeted object-mislabeling. We develop the TOG family of attacks, which on one hand show the feasibility of attacking one-phase regression-based and two-phase proposal-based detectors using the same attack framework, and on the other hand provide a broader coverage of vulnerabilities for analyzing and understanding object detection through an adversarial lens. (2) Our evaluation framework provides two main building blocks: the attack module, which incorporates the state-of-the-art attack algorithms, and the evaluation module, which includes a set of quantitative metrics to measure, compare and analyze different attack algorithms in terms of adversarial effectiveness and costs, and attack transferability. We define cross-model transferability in terms of both algorithm and backbone of the detectors and introduce cross-resolution transferability to enrich our analysis on adversarial robustness of deep object detection models. (3) We conduct comprehensive experimental analysis on six object detectors from three dominant families of object detection algorithms (YOLOv3, SSD, and Faster R-CNN), with four representative attack methods: DAG~\cite{xie2017adversarial}, RAP~\cite{li2018robust}, UEA~\cite{wei2018transferable}, and TOG~\cite{chow2020tog}, on two benchmark datasets: PASCAL VOC~\cite{everingham2015pascal} and MS COCO~\cite{lin2014microsoft}. Our experimental results further demonstrate the utility of the proposed framework as a methodical benchmark platform for evaluating adversarial robustness of deep object detectors, and assessing the security risks and the attack resilience of deep object detectors to be deployed in real-world applications.

\vspace{-8pt}\section{Proposed Framework - Attack Module}
\begin{figure}[t]
	\centering
	\includegraphics[width=0.84\textwidth]{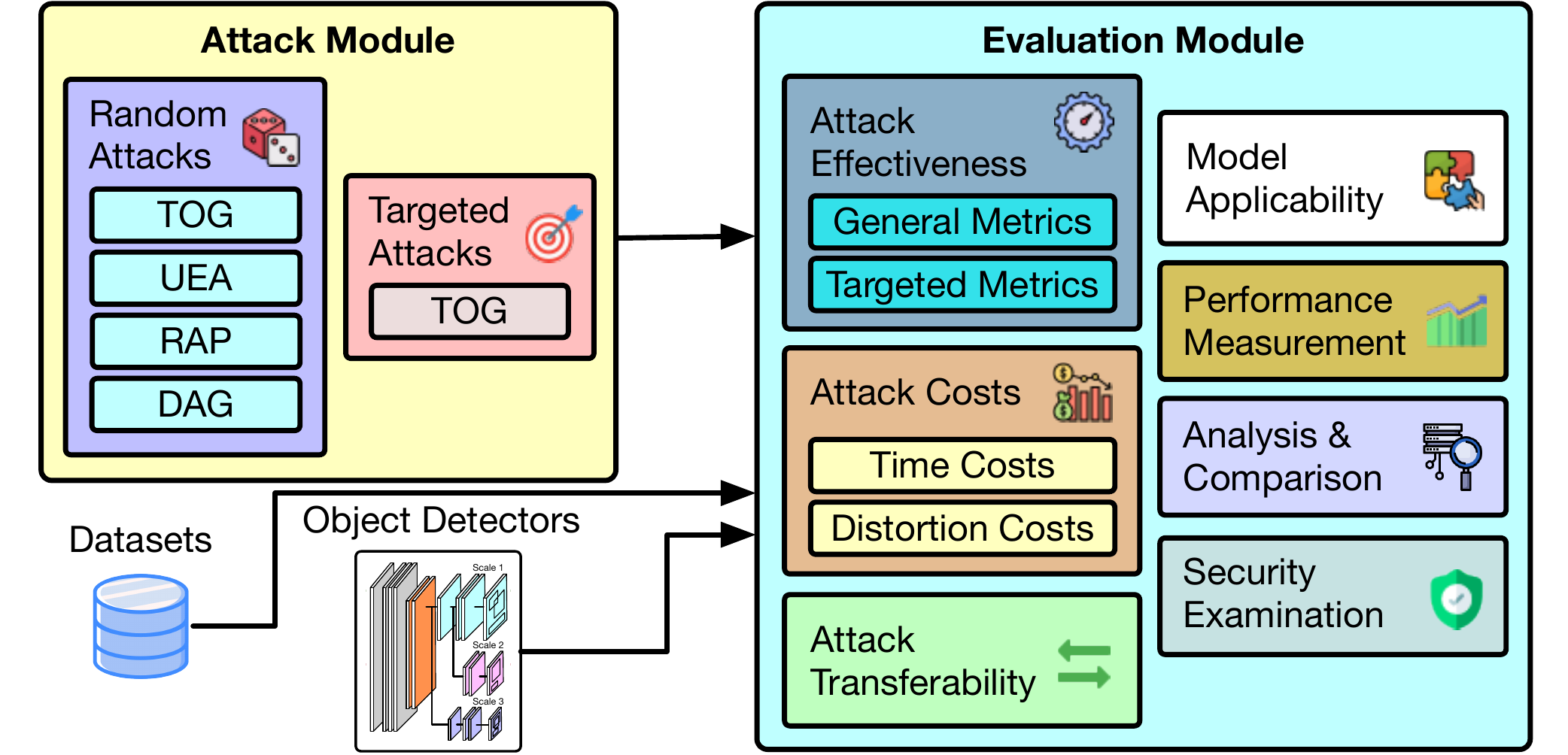}
	\vspace{-4pt}
	\caption{The overview of the evaluation framework.}
	\vspace{-12pt}
	\label{fig:eval-system-overview}
\end{figure}

\vspace{-6pt}~\label{sec:attack-module} 
Figure~\ref{fig:eval-system-overview} gives an overview of the proposed framework. This section is dedicated to the attack module, a collection of attack algorithms for comparisons and analysis. We first give an algorithmic overview of deep object detection algorithms and adversarial attacks. Then, we provide the formal analysis on the four state-of-the-art attack algorithms (TOG~\cite{chow2020tog}, UEA~\cite{wei2018transferable}, RAP~\cite{li2018robust}, and DAG~\cite{xie2017adversarial}).

\vspace{-20pt}\subsection{DNN-based Object Detection and Adversarial Attacks}\vspace{-4pt}\label{sec:overview-od}
DNN-based object detection is a multi-task learning problem, aiming to minimize the prediction error of (1) object existence, (2) bounding boxes, and (3) class labels of detected objects. Given an input image $\boldsymbol{x}$ with resolution $(H\times W)$, a $K$-class object detector $f_{\boldsymbol{\theta}}$, parameterized by $\boldsymbol{\theta}$, generates a large number of $S$ candidate objects  $\{\hat{\boldsymbol{o}}_1,\dots,\hat{\boldsymbol{o}}_S\}$ where $\hat{\boldsymbol{o}}_i=(\hat{b}^x_i, \hat{b}^y_i, \hat{b}^w_i, \hat{b}^h_i, \hat{C}_i, \hat{\boldsymbol{p}}_i)$ represents a candidate centered at coordinates $(\hat{b}^x_i, \hat{b}^y_i)$ having a dimension $(\hat{b}^w_i, \hat{b}^h_i)$ with an objectness probability of $\hat{C}_i\in[0,1]$ to be a real object, and a $K$-class probability vector $\hat{\boldsymbol{p}}_i=(\hat{p}^1_i, \hat{p}^2_i,\dots,\hat{p}^K_i)$. This is often done by dividing the input into mesh grids in different scales (resolutions). Each grid cell is responsible for locating objects centered at the cell. The final detection results $\hat{\boldsymbol{\mathcal{O}}}$ are obtained by applying confidence thresholding to remove candidates with low prediction confidence and non-maximum suppression to exclude those with high overlapping.

To train a deep object detection neural network, every ground-truth object in a training sample $\tilde{\boldsymbol{x}}$ is assigned to one of the $S$ candidates according to their center coordinates. Let $\boldsymbol{\mathcal{O}}$ be the set of ground-truth objects of $\tilde{\boldsymbol{x}}$. The object detector can be trained by optimizing the following multi-task learning objective:
\vspace{-8pt}
\begin{equation}\small\label{eq:detector-loss}
\mathcal{L}(\tilde{\boldsymbol{\mathcal{D}}}; \boldsymbol{\theta})=\mathbbm{E}_{(\tilde{\boldsymbol{x}}, \boldsymbol{\mathcal{O}})\in\tilde{\boldsymbol{\mathcal{D}}}}[\mathcal{L}_\text{obj}(\tilde{\boldsymbol{x}},\boldsymbol{\mathcal{O}}; \boldsymbol{\theta})+\mathcal{L}_\text{bbox}(\tilde{\boldsymbol{x}},\boldsymbol{\mathcal{O}}; \boldsymbol{\theta})+\mathcal{L}_\text{class}(\tilde{\boldsymbol{x}},\boldsymbol{\mathcal{O}}; \boldsymbol{\theta})]
\vspace{-1pt}
\end{equation}
where $\tilde{\boldsymbol{\mathcal{D}}}$ is the training set, $\mathcal{L}_\text{obj}$, $\mathcal{L}_\text{bbox}$, and $\mathcal{L}_\text{class}$ represent the loss function of the three prediction tasks: object existence (objectness), object localization (bounding box), and object class label respectively. In the rest of this paper, we use $\boldsymbol{\mathcal{O}}$ and $\hat{\boldsymbol{\mathcal{O}}}$ to distinguish between ground-truth and predicted detection, and we only specify the argument (e.g., $\boldsymbol{\mathcal{O}}(\boldsymbol{x})$) to emphasize the input if necessary. 

An adversarial example $\boldsymbol{x}'$ is generated by perturbing a benign input $\boldsymbol{x}$ sent to the victim detector, aiming to fool the victim to misdetect randomly or purposefully. The generation process can be conceptually formulated as
\vspace{-4pt}
\begin{equation}\small\label{eq:adv-process}
\min||\boldsymbol{x}' - \boldsymbol{x}||_{p} \quad s.t.\;\hat{\boldsymbol{\mathcal{O}}}(\boldsymbol{x}')\neq\hat{\boldsymbol{\mathcal{O}}}(\boldsymbol{x}), \hat{\boldsymbol{\mathcal{O}}}(\boldsymbol{x}')={\boldsymbol{\mathcal{O}}}^*(\boldsymbol{x})
\vspace{-4pt}
\end{equation}
where $p$ is the distance metric and ${\boldsymbol{\mathcal{O}}}^*(\boldsymbol{x})$ is the incorrect detection. Popular choices for the distance metric include the $L_\infty$ norm,  denoting the maximum change to any pixel, the $L_2$ norm, computing the Euclidean distance, and the $L_0$ norm, measuring the number of the pixels that are changed.

Although adversarial attacks on object detection systems are more sophisticated, adopting different formulations, they generally exploit gradients derived from one or multiple losses in Equation~\ref{eq:detector-loss} (i.e., $\mathcal{L}_\text{obj}$, $\mathcal{L}_\text{bbox}$, and $\mathcal{L}_\text{class}$). This allows the attack algorithm to meticulously inject perturbations to the input image, such that the tiny changes in input will be amplified throughout the forward propagation of the victim detector, and become large enough to alter one or more types of prediction results (i.e., object existence, bounding box, and class probability), depending on the composition of gradients. We analyze below the four representative attack algorithms on object detection systems, understanding their properties and demystifying their working principles.

\vspace{-8pt}\subsection{TOG: Targeted Objectness Gradient Attacks}\vspace{-4pt}\label{sec:tog}
We develop the TOG family of attacks~\cite{chow2020tog} based on an iterative gradient approach to obtain the malicious perturbation fooling the victim detector to give the desired erroneous detection. With a proper setting of the designated detection $\boldsymbol{\mathcal{O}}^*(\boldsymbol{x})$ and the attack loss $\mathcal{L}^*$, TOG can be generally formulated as:
\vspace{-4pt}
\begin{equation}\small
\boldsymbol{x}'_{t+1}=\prod\limits_{\boldsymbol{x}, \epsilon}\big[\boldsymbol{x}'_t - \alpha_\text{TOG}\Gamma\big({\nabla_{\boldsymbol{x}'_t} \mathcal{L}^*(\boldsymbol{x}'_t, \boldsymbol{\mathcal{O}}^*(\boldsymbol{x}); \boldsymbol{\boldsymbol{\theta}})}\big)\big]
\vspace{-7pt}
\end{equation}
where $\boldsymbol{x}'_t$ is the adversarial example at the $t$-th iteration, $\prod_{\boldsymbol{x},\epsilon}[\cdot]$ is the projection onto a hypersphere with a radius $\epsilon$ centered at $\boldsymbol{x}$ in $L_p$ norm, $\Gamma$ is a sign function, and $\alpha_\text{TOG}$ is the attack learning rate. With this formulation, TOG allows adversaries to specify the effect imposed on victim's detection accuracy and correctness, including untargeted random attacks and three types of targeted specificity attacks: object-vanishing, object-fabrication, and targeted object-mislabeling. 

\textit{Untargeted} attacks fool the victim detector to \emph{randomly} misdetect without targeting at any specific object. This class of attacks succeeds if the adversarial example fools the victim detector to give incorrect result of any form, such as having objects vanished, fabricated, or mislabeled randomly. TOG exploits gradients from both $\mathcal{L}_\text{obj}$, $\mathcal{L}_\text{bbox}$, and $\mathcal{L}_\text{class}$ and formulates the attack to be
\vspace{-4pt}
\begin{equation}\small
\boldsymbol{x}'_{t+1}=\prod\limits_{\boldsymbol{x}, \epsilon}\big[\boldsymbol{x}'_t + \alpha_\text{TOG}\Gamma\big({\nabla_{\boldsymbol{x}'_t} \mathcal{L}(\boldsymbol{x}'_t, \boldsymbol{\mathcal{O}}(\boldsymbol{x});\boldsymbol{\boldsymbol{\theta}})}\big)\big].
\vspace{-7pt}
\end{equation}
As shown in the $2$nd column in Table~\ref{tab:effect-examples}, the victim detector cannot identify any correct objects that were detected on benign inputs ($1$st column) but the exact effect varies across input images and attack algorithms. 

\textit{Object-vanishing} attacks \emph{consistently} disable the victim detector to locate and recognize any object. 
TOG-vanishing utilizes gradients from $\mathcal{L}_\text{obj}$ as it dominates the decision on object existences and formulates the attack as follows:
\vspace{-6pt}
\begin{equation}\small
\boldsymbol{x}'_{t+1}=\prod\limits_{\boldsymbol{x}, \epsilon}\big[\boldsymbol{x}'_t - \alpha_\text{TOG}\Gamma\big({\nabla_{\boldsymbol{x}'_t} \mathcal{L}_\text{obj}(\boldsymbol{x}'_t,\O; \boldsymbol{\boldsymbol{\theta}})}\big)\big]
\vspace{-4pt}
\end{equation}
By targeting specifically at object-vanishing, this attack if successful will make the victim detector fail to detect any object as shown in the $3$rd column in Table~\ref{tab:effect-examples} where no object is detected in both examples.

\textit{Object-fabrication} attacks \emph{consistently} fool the victim to mistakenly recognize false objects. 
TOG-fabrication leverages gradients from $\mathcal{L}_\text{obj}$ with formulation:
\vspace{-4pt}
\begin{equation}\small
\boldsymbol{x}'_{t+1}=\prod\limits_{\boldsymbol{x}, \epsilon}\big[\boldsymbol{x}'_t + \alpha_\text{TOG}\Gamma\big({\nabla_{\boldsymbol{x}'_t} \mathcal{L}_\text{obj}(\boldsymbol{x}'_t,\O; \boldsymbol{\boldsymbol{\theta}})}\big)\big].
\vspace{-6pt}
\end{equation}
This attack makes the victim to drastically increase the number of detected objects by introducing fake objects, as illustrated in the $4$th column in Table~\ref{tab:effect-examples}.

\textit{Targeted object-mislabeling} attacks \emph{consistently} cause the victim detector to misclassify the objects detected on the input image by replacing their source class label with the maliciously chosen target class label, while maintaining the same set of correct bounding boxes. By focusing on the classification loss (i.e., $\mathcal{L}_\text{class}$) and keeping the gradients of the other two parts unchanged, TOG-mislabeling assigns the target class label to each object in $\boldsymbol{\mathcal{O}}(\boldsymbol{x})$ to form $\boldsymbol{\mathcal{O}}^*(\boldsymbol{x})$ and generate adversarial examples with
\vspace{-6pt}
\begin{equation}\small
\boldsymbol{x}'_{t+1}=\prod\limits_{\boldsymbol{x}, \epsilon}\big[\boldsymbol{x}'_t - \alpha_\text{TOG}\Gamma\big({\nabla_{\boldsymbol{x}'_t} \mathcal{L}(\boldsymbol{x}'_t,\boldsymbol{\mathcal{O}}^*(\boldsymbol{x}); \boldsymbol{\boldsymbol{\theta}})}\big)\big].
\vspace{-8pt}
\end{equation}
For instance, the object-mislabeling attack in the $5$th column in Table~\ref{tab:effect-examples} is configured to fool the victim to mislabel any stop sign as an umbrella. Note that the person (top) and the car (bottom) can still be detected under this attack as they are not the objects of attack interest and only stop signs will be mislabeled.

As TOG does not attack a special structure (e.g., RPN) in an object detector, it is applicable to both one-phase and two-phase techniques. Inspired by the universal perturbations to attack image classifiers~\cite{moosavi2017universal}, TOG also develops universal perturbations to attack deep object detectors in terms of object-vanishing or object-fabrication attack~\cite{chow2020tog}. By training the universal perturbation offline on a training set and a victim detector, the universal perturbation can be applied during the online detection phase to any input sent to the victim. 

\vspace{-8pt}\subsection{DAG: Dense Adversary Generation}\vspace{-4pt}
DAG~\cite{xie2017adversarial} is an untargeted random attack and begins with manually assigning the IOU threshold to $0.90$ in non-maximum suppression (NMS) in the RPN of a given two-phase model. This attack setting requires one proposal region to be highly overlapped ($>90\%$) with the other proposal region in order to be pruned. Hence, a large amount of proposal regions remain unpruned. After the refinement by the subsequent network for bounding box and class label prediction, DAG assigns a randomly selected label for each proposal region and then performs the iterative gradient attack to misclassify the proposals with the following formulation:
\vspace{-6pt}
\begin{equation}\small\label{eq:dag}
\boldsymbol{r}_t=\nabla_{\boldsymbol{x}'_t}\sum_{j=1}^Jz_j[p^c_j-p^{c'}_j],\;\; \boldsymbol{x}'_{t+1}=\boldsymbol{x}'_t - \frac{\alpha_\text{DAG}}{\vert\vert\boldsymbol{r}_t\vert\vert_\infty}\boldsymbol{r}_t
\vspace{-8pt}
\end{equation}
where $z_j=1$ if the $j$-th proposal on $\boldsymbol{x}'_t$ from RPN is foreground and $0$ otherwise, $p^c_j$ and $p^{c'}_j$ are the prediction confidence of the correct class $c$ and randomly selected incorrect class $c'$ of the $j$-th proposal and $\alpha_\text{DAG}$ is the attack learning rate. This is equivalent to exploiting gradients derived from the classification loss $\mathcal{L}_\text{class}$. As DAG requires to manipulate the RPN to generate a large number of proposals, it can only be directly applicable to two-phase detection models.

\vspace{-8pt}\subsection{RAP: Robust Adversarial Perturbation }\vspace{-4pt}
RAP~\cite{li2018robust} is an untargeted random attack and focuses on collapsing the function of the RPN in two-phase algorithms. It exploits the composite gradients from (i) the objectness loss, i.e., $\mathcal{L}_\text{obj}$, that fools the RPN to not returning foreground objects, and (ii) the localization loss, i.e., $\mathcal{L}_\text{bbox}$, that causes the bounding box estimation to be incorrect even if foreground objects are proposed:
\vspace{-8pt}
\begin{equation}\small
\boldsymbol{r}_t=\nabla_{\boldsymbol{x}'_t}\sum_{j=1}^{J}z_j[\log(\hat{C}_j)+\ell_{\text{SE}}(\hat{\boldsymbol{b}_j}, \boldsymbol{\tau})], \;\; \boldsymbol{x}'_{t+1}=\boldsymbol{x}'_t - \frac{\alpha_\text{RAP}}{\vert\vert\boldsymbol{r}_t\vert\vert_2}\boldsymbol{r}_t
\vspace{-8pt}
\end{equation}
where $\ell_\text{SE}$ is the squared error, $\hat{\boldsymbol{b}}_j$ and $\boldsymbol{\tau}$ are quadruples of the proposed bounding box and large offsets respectively, and $\alpha_\text{RAP}$ is the attack learning rate.

\vspace{-8pt}\subsection{UEA: Unified and Efficient Adversary}\vspace{-4pt}\label{sec:uea}
UEA~\cite{wei2018transferable} is an untargeted random attack. It trains a conditional generative adversarial network (GAN)~\cite{goodfellow2014generative} to craft adversarial examples. In deep object detectors, the backbone network plays an important role in feature extraction for region proposals in two-phase algorithms or object recognition in one-phase techniques. In practice, it is often one of the popular architectures (e.g., VGG16) that perform well in large-scale image classification and is pretrained with the ImageNet dataset for transfer learning. UEA designs a multi-scale attention feature loss, encouraging the GAN to create adversarial examples that can corrupt the feature map extracted by the backbone network in the victim detector:
\vspace{-8pt}
\begin{equation}\small\label{eq:uea}
\mathcal{L}^\text{Fea}_\text{UEA}=\mathbbm{E}_{(\tilde{\boldsymbol{x}}, \boldsymbol{\mathcal{O}})\in\boldsymbol{\mathcal{D}}}[\sum_{m=1}^M\vert\vert\mathbf{A}_m\circ(\boldsymbol{\tilde{x}}_m-\mathbf{R}_m)\vert\vert_2]
\vspace{-8pt}
\end{equation}
where $\boldsymbol{\tilde{x}}_m$ is the extracted feature map of the training example $\boldsymbol{\tilde{x}}$ in the $m$-th layer of the backbone network, $\mathbf{R}_m$ is a randomly predefined feature map, and $\mathbf{A}_m$ is the attention weight computed based on the proposal regions from the RPN. Whenever another detector is equipped with the same backbone, the adversarial examples are likely to be effective. Equation~\ref{eq:uea} is jointly optimized with the DAG formulation (Equation~\ref{eq:dag}), requiring the manipulation of the RPN. Hence, it is unable to directly attack one-phase algorithms.

\vspace{-8pt}\section{Proposed Framework - Evaluation Module}\vspace{-8pt}\label{sec:evaluation-module}
The evaluation module is the second building block of the proposed framework (Figure~\ref{fig:eval-system-overview}), providing experimental testbed to measure, evaluate and analyze attacks and adversarial robustness of an object detector from four perspectives.

\vspace{-8pt}\subsection{Attack Effectiveness}\vspace{-4pt}
\textbf{mean Average Precision (mAP).} The interpolated average precision (AP) has been used by major object detection competitions~\cite{everingham2015pascal,lin2014microsoft}. For a given class, the precision/recall curve is computed from the detector's output, ranked by the detected confidence. The AP summarizes the shape of the precision/recall curve by taking the mean precision at a set of equally spaced recall levels. Then, the mean Average Precision (mAP) that quantifies the overall detection quality of a detector is computed by taking the mean of APs of all classes. The general attack performance can be analyzed on two sets of mAP (or AP), one on benign examples and another on adversarial examples. A low adversarial mAP implies the power of the attack but reveals the vulnerability of the victim model.

\textbf{Attack Success Rate (ASR).} In addition to comparing mAPs to reveal the impact on overall performance of the victim, we further define the attack success rate (ASR) for each targeted specificity attack, to capture their capability to fool the victim to misbehave with the designated effect (e.g., object-vanishing).

For object-vanishing attacks, we define the ASR as the proportion of objects detected on benign examples that are not covered by any objects detected on their adversarial counterparts:
\vspace{-6pt}
\begin{equation}\small
\text{ASR}=\frac{\sum_{\boldsymbol{x}\in\boldsymbol{\mathcal{D}}}\sum_{\hat{\boldsymbol{o}}\in\hat{\boldsymbol{\mathcal{O}}}(\boldsymbol{x})}\mathbbm{1}[\neg\exists\hat{\boldsymbol{o}}'\in\hat{\boldsymbol{\mathcal{O}}}(\boldsymbol{x}')(\text{IOU}(\hat{\boldsymbol{o}}_{\text{[bbox]}}, \hat{\boldsymbol{o}}_{\text{[bbox]}}')\geq t_{\text{IOU}})]}{\sum_{\boldsymbol{x}\in\boldsymbol{\mathcal{D}}}\vert\vert\hat{\boldsymbol{\mathcal{O}}}(\boldsymbol{x})\vert\vert},
\vspace{-4pt}
\end{equation} 
where $\mathbbm{1}[\text{condition}] = 1$ if the condition is met and $0$ otherwise, IOU$(\hat{\boldsymbol{o}}_{\text{[bbox]}}, \hat{\boldsymbol{o}}_{\text{[bbox]}}')$ computes the intersection over union of the two bounding boxes $\hat{\boldsymbol{o}}_{\text{[bbox]}}$ and $\hat{\boldsymbol{o}}'_{\text{[bbox]}}$, and $t_\text{IOU}$ is a predefined threshold controlling the amount of overlapping required for two bounding boxes to be considered as referring to the same entity. 

For object-fabrication attacks, the ASR is defined as the proportion of test examples where additional false objects are mistakenly detected by the victim detector under attacks:
\vspace{-4pt}
\begin{equation}\small
\text{ASR}= \frac{1}{\vert\vert\boldsymbol{\mathcal{D}}\vert\vert}\sum_{\boldsymbol{x}\in\boldsymbol{\mathcal{D}}}\mathbbm{1}[\vert\vert\hat{\boldsymbol{\mathcal{O}}}(\boldsymbol{x}')\vert\vert > \vert\vert\hat{\boldsymbol{\mathcal{O}}}(\boldsymbol{x})\vert\vert].
\vspace{-4pt}
\end{equation}

For object-mislabeling attacks, we define the ASR to be the proportion of objects detected on benign examples that are mislabeled as the target label by the victim detector on their adversarial counterparts:
\vspace{-4pt}
\begin{equation}\small\label{eq:asr-mislabeling}
\text{ASR}=\frac{\sum_{\boldsymbol{x}\in\boldsymbol{\mathcal{D}}}\sum_{\hat{\boldsymbol{o}}\in\hat{\boldsymbol{\mathcal{O}}}(\boldsymbol{x})}\mathbbm{1}[\exists\hat{\boldsymbol{o}}'\in\hat{\boldsymbol{\mathcal{O}}}(\boldsymbol{x}')(\text{IOU}(\hat{\boldsymbol{o}}_{\text{[bbox]}},\hat{\boldsymbol{o}}'_\text{[bbox]})\geq t_\text{IOU}\land \hat{\boldsymbol{o}}'_\text{[class]}=\mathcal{T}(\hat{\boldsymbol{o}}_\text{[class]}))]}{\sum_{\boldsymbol{x}\in\boldsymbol{\mathcal{D}}}\vert\vert\hat{\boldsymbol{\mathcal{O}}}(\boldsymbol{x})\vert\vert}
\vspace{-4pt}
\end{equation}
where $\mathcal{T}(\hat{\boldsymbol{o}}_\text{[class]})$ is a mapping from a source class to a target class. Under this setting, we consider the attack succeeds only if it (i) does not alter the bounding box significantly and (ii) fools the detector to give a designated wrong label. 

\vspace{-8pt}\subsection{Attack Cost}\vspace{-4pt}
\textbf{Time Cost. } We measure time cost using two metrics:
(i) the attack time, which measures the additional time introduced by the attack, excluding the inference of the victim detector to obtain the final detection results; and
(ii) the total time cost, which considers both attack time and (benign) detection time. 

\textbf{Distortion Cost. } Remaining human-imperceptible is an important factor in adversarial attacks as significant distortion naturally mislead a deep learning model to misbehave. A robust object detection model should be resilient against adversarial examples that are visually identical to their benign counterparts. 

${L_0}$, ${L_2}$, and ${L_\infty}$ distances have been popularly used in adversarial learning. They are used as a constraint to limit the maximum perturbation introducible to the benign example. Note that a low $L_p$ distance means a high imperceptibility.

Structural Similarity (SSIM) has become an important metric to quantify the similarly between two images in computer vision:
\vspace{-6pt}
\begin{equation}\small
\text{SSIM}(\boldsymbol{x}, \boldsymbol{x}')=\sum_{i=1}^{I}\frac{(2\mu_{\boldsymbol{x}[i]}\mu_{\boldsymbol{x}'[i]}+\kappa_1)(2\sigma_{\boldsymbol{x}[i]\boldsymbol{x}'[i]}+\kappa_2)}{(\mu^2_{\boldsymbol{x}[i]}+\mu^2_{\boldsymbol{x}'[i]}+\kappa_1)(\sigma^2_{\boldsymbol{x}[i]}+\sigma^2_{\boldsymbol{x}'[i]}+\kappa_2)}
\vspace{-4pt}
\end{equation}
where $\boldsymbol{x}[i]$ denotes the $i$-th channel of image $\boldsymbol{x}$, $\mu_{\boldsymbol{x}}$ and $\sigma_{\boldsymbol{x}}$ are the average and variance of $\boldsymbol{x}$ respectively, $\sigma_{\boldsymbol{x}\boldsymbol{x}'}$ is the covariance of $\boldsymbol{x}$ and $\boldsymbol{x}$', and $\kappa_1$ and $\kappa_2$ are two variables for numerical stability. It has a range from $-1.00$ (the least similar) to $1.00$ (the most similar) and is considered to be more consistent to human visual perception than $L_p$ distances. As attacks optimize different $L_p$ distances, SSIM offers an objective comparison on the imperceptibility of adversarial examples.

\vspace{-8pt}\subsection{Attack Transferability}\vspace{-4pt}
All adversarial attacks on deep object detectors are white-box attacks as they require model weights to optimize the generation of adversarial perturbation against a victim detector. The transferability of adversarial examples generated against one victim detector can be utilized to launch black-box attacks to other detectors, in a similar way as the transferability of adversarial examples to attack different image classifiers~\cite{papernot2016transferability}. For object detection, we propose to study not only the cross-model transferability, but also the cross-resolution transferability.

\emph{Cross-model transferability} in object detection can be further broken down into (i) cross-algorithm transferability that the source and the target models use different detection algorithms and (ii) cross-backbone transferability that examines the transferability between different backbones of the same detection algorithm and between different detection algorithms with the same backbone.

\emph{Cross-resolution transferability} covers a characteristic unique to those object detection algorithms (e.g., YOLO and Faster R-CNN) that allow variable input resolutions. In contrast to image classification networks where the resolution of the input image is fixed due to the fully-connected layer for the final softmax, for object detection, increasing input resolution can generate more candidate objects with a potentially better detection quality with the cost of slowing down the detection. The cross-resolution transferability reveals whether the adversarial examples generated by an attack algorithm on a source resolution can be robust and survive under resizing and interpolation to the target resolution.

\vspace{-8pt}\subsection{Model Applicability}\vspace{-4pt}
From a macroscopic perspective, all object detection systems take an input image and output a set of detected objects. They may appear to be similar, but their internal learn-to-detect mechanisms can be very different. Some existing attacks are designed by exploiting the vulnerability of a particular structure, e.g., the region proposal network (RPN) in Faster R-CNN detectors. Hence, not all attack techniques are universally applicable. RAP~\cite{li2018robust} is an example, which perturbs the benign image to disable the functionality of the RPN in two-phase algorithms and cannot be used on one-phase detectors where no RPN is used. We also leverage model-applicability as an evaluation aspect on attack algorithms. 

\vspace{-8pt}\section{Experimental Analysis}\vspace{-4pt}\label{sec:experiment}
Extensive experiments are conducted on two benchmark datasets: PASCAL VOC~\cite{everingham2015pascal} and MS COCO~\cite{lin2014microsoft}. All results are based on the entire test set, and we preprocess images by padding to preserve the aspect ratio of objects. We consider six models from three dominant detection algorithms. YOLOv3-D and YOLOv3-M are two YOLOv3~\cite{redmon2018yolov3} models with a Darknet53 and a MobileNetV1 backbone respectively. For SSD~\cite{liu2016ssd}, we have SSD300 and SSD512 corresponding to two models with different input resolutions. Finally, FRCNN denotes the Faster R-CNN~\cite{ren2015faster} model. As experimental results on COCO are highly similar to VOC, we provide only YOLOv3-D on COCO due to the space constraint. We provide more experimental configuration details in Appendix A. 

\vspace{-12pt}\subsection{Untargeted Random Attacks}\vspace{-4pt}
This section reports the set of experiments to compare the four attack algorithms: TOG, UEA, RAP, and DAG in terms of effectiveness and time cost of untargeted attacks.
\begin{table}[t]\renewcommand{\arraystretch}{0.8}
	\centering\scriptsize
	\begin{tabular}{ccccc}
		\textbf{No Attack} & \textbf{TOG} & \textbf{UEA} & \textbf{RAP} & \textbf{DAG} \\ 
		\includegraphics[scale=0.105]{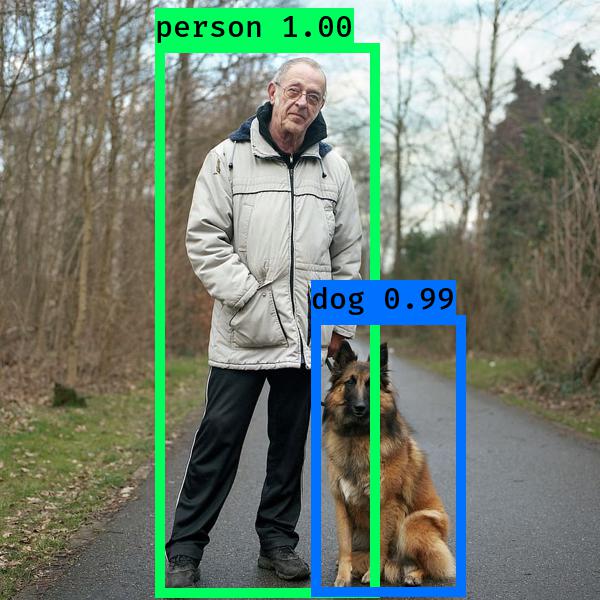} & \includegraphics[scale=0.105]{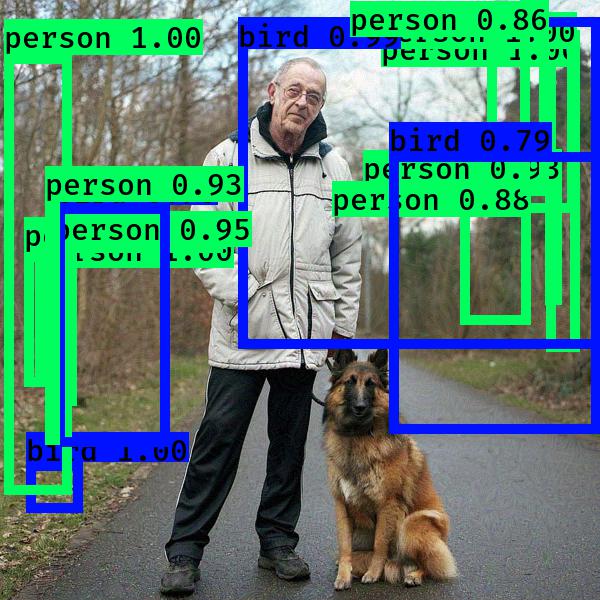} & \includegraphics[scale=0.1051]{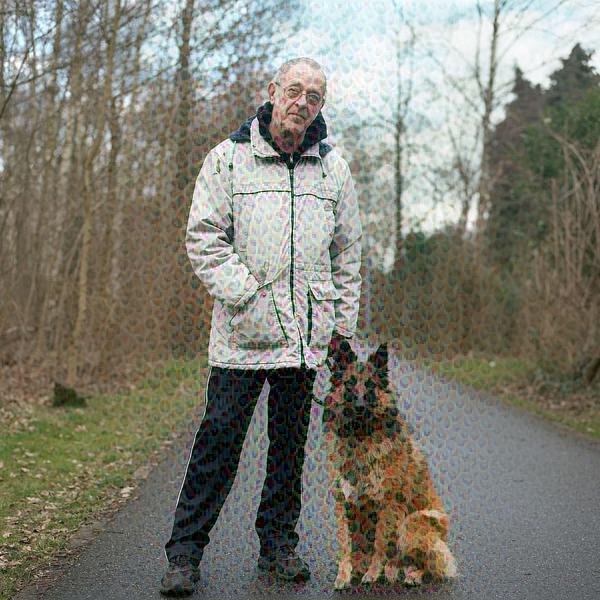} & \includegraphics[scale=0.105]{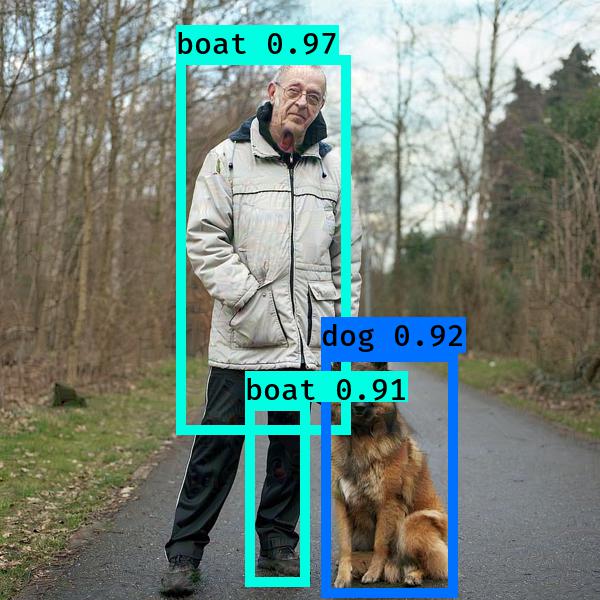} & \includegraphics[scale=0.105]{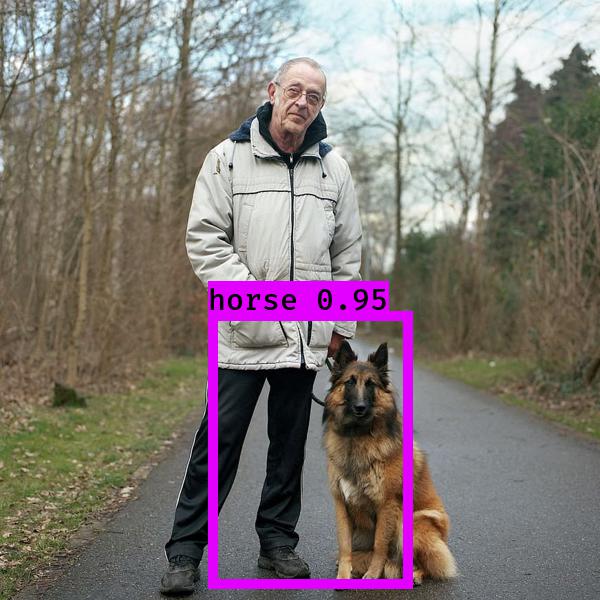} \\
		\includegraphics[scale=0.105]{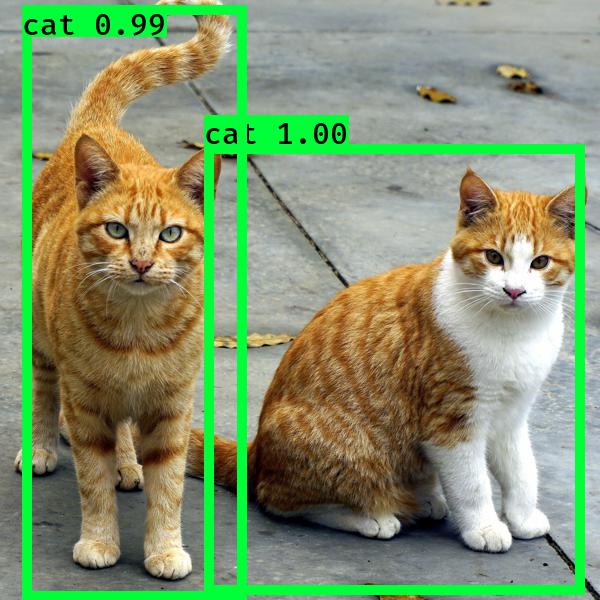} & \includegraphics[scale=0.105]{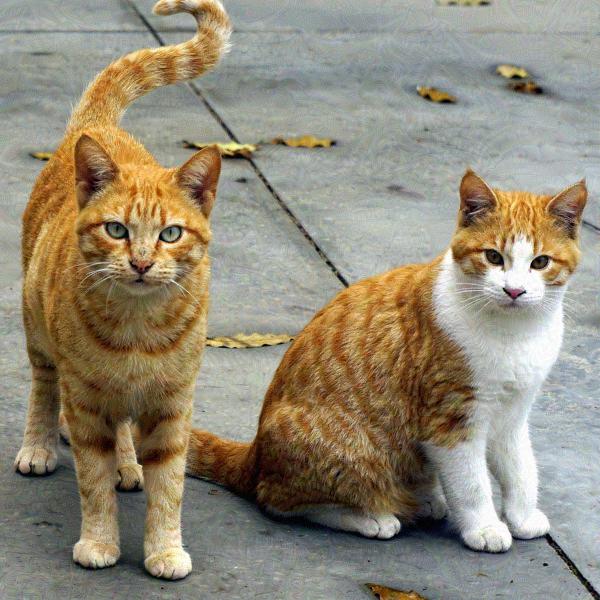} & \includegraphics[scale=0.105]{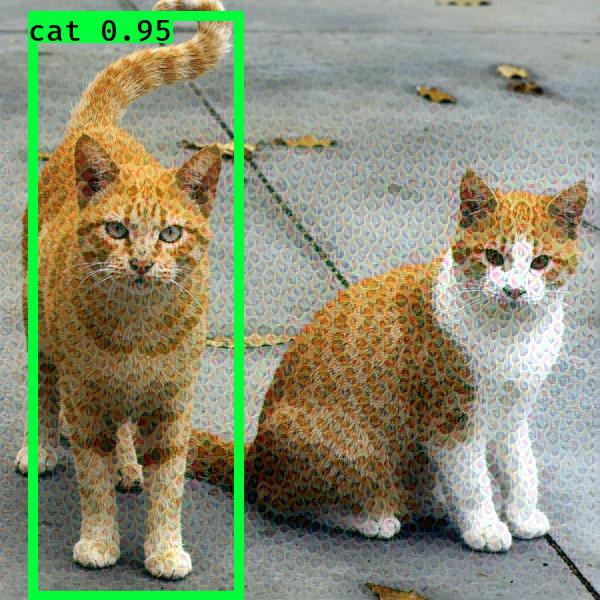} & \includegraphics[scale=0.105]{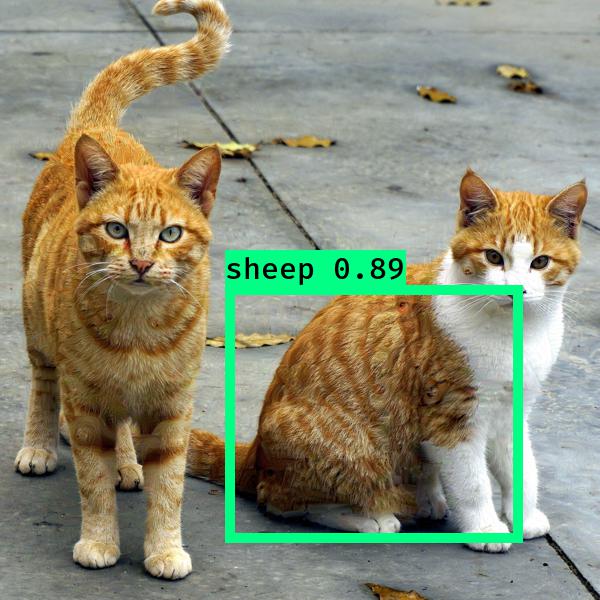} & \includegraphics[scale=0.105]{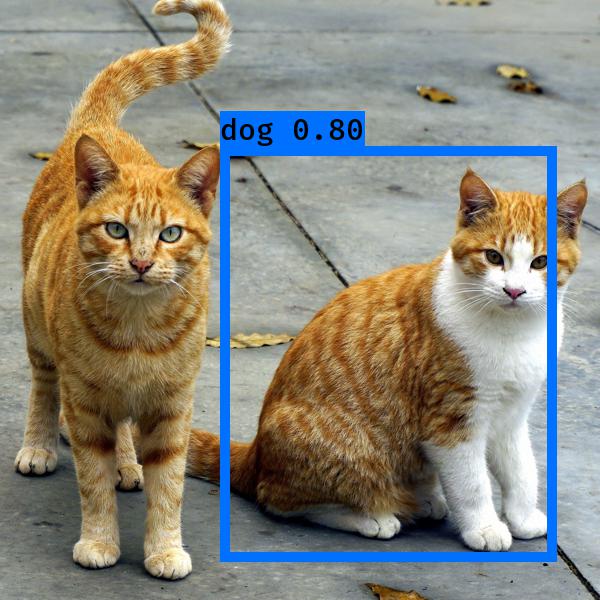} \\
		\includegraphics[scale=0.105]{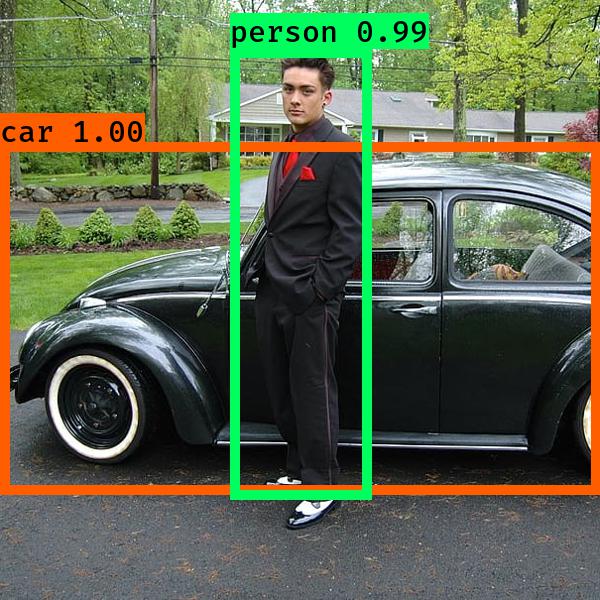} & \includegraphics[scale=0.105]{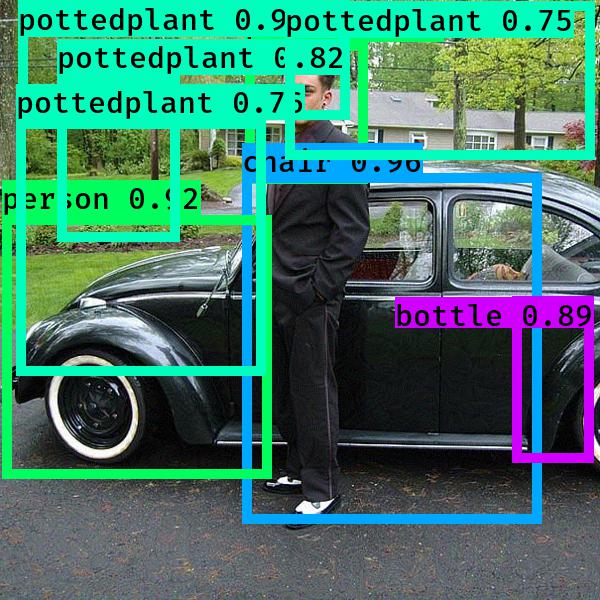} & \includegraphics[scale=0.105]{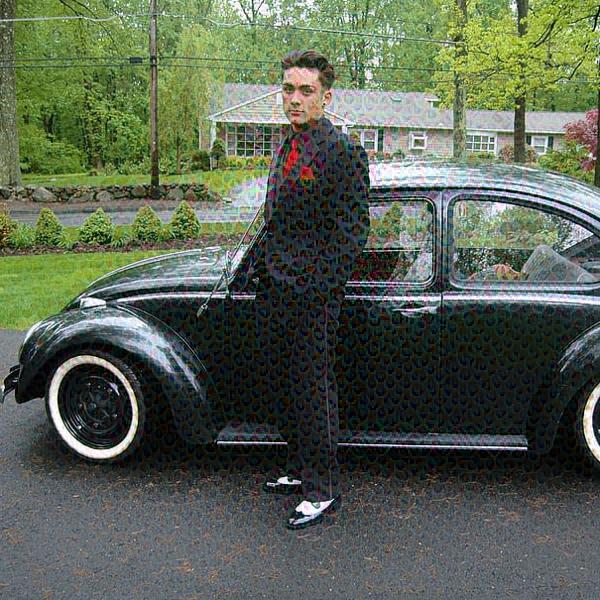} & \includegraphics[scale=0.105]{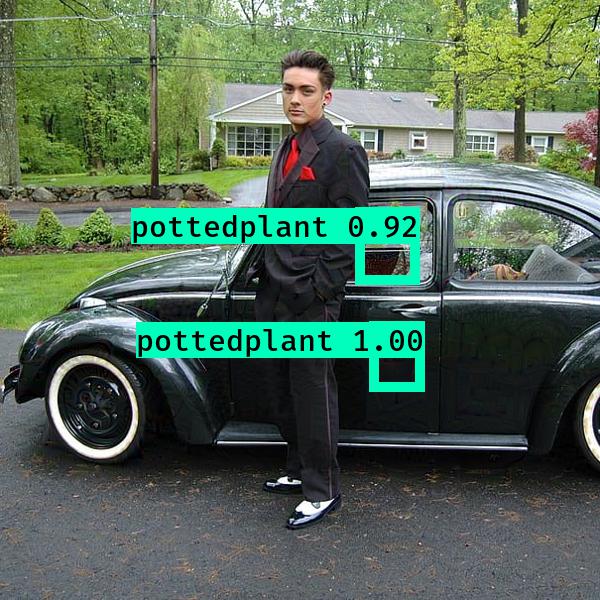} & \includegraphics[scale=0.105]{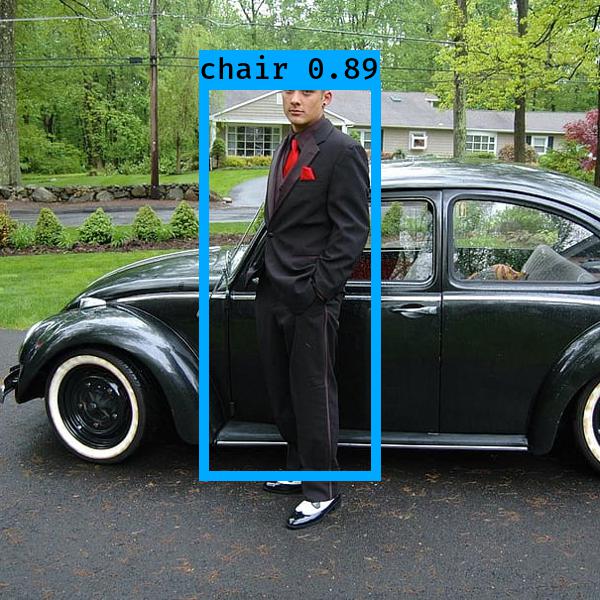} \\
		\includegraphics[scale=0.105]{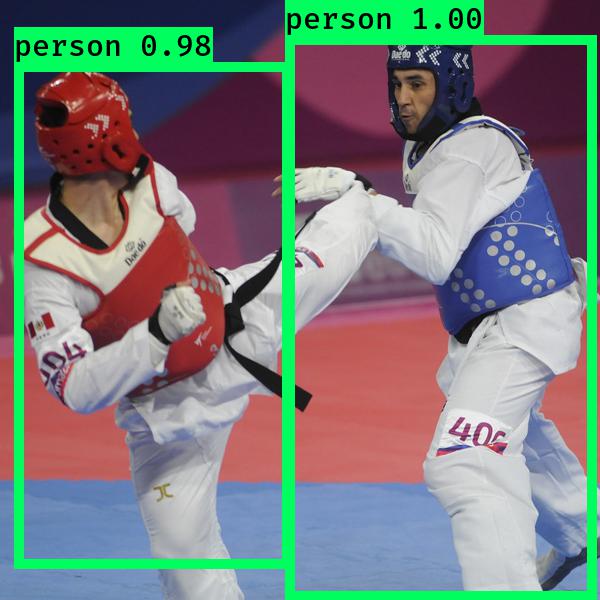} & \includegraphics[scale=0.105]{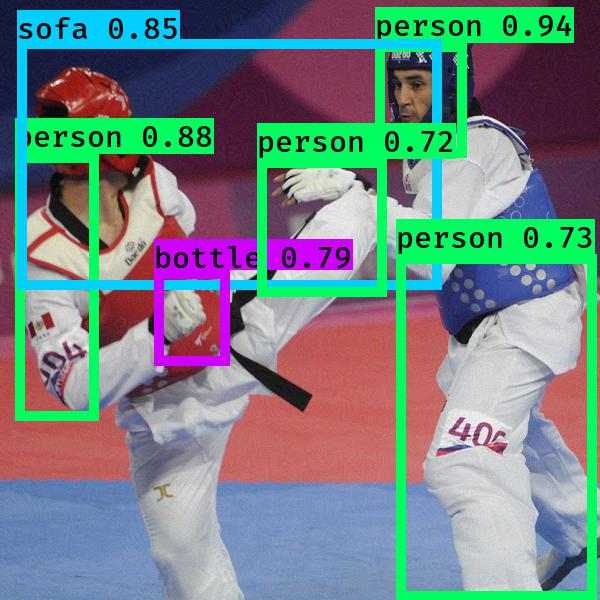} & \includegraphics[scale=0.105]{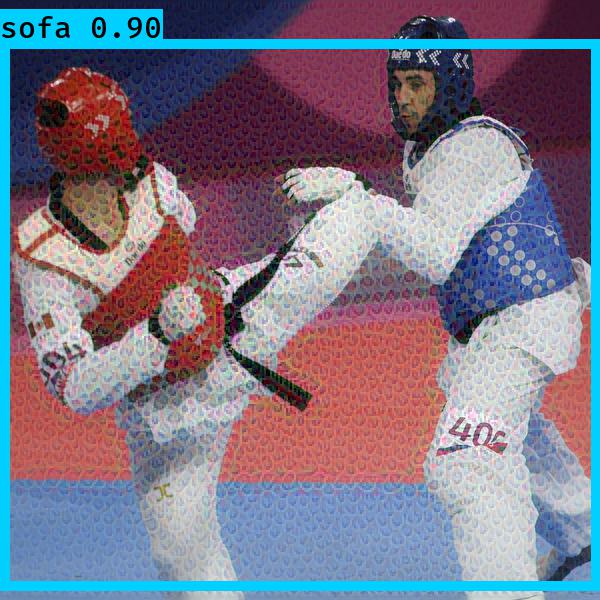} & \includegraphics[scale=0.105]{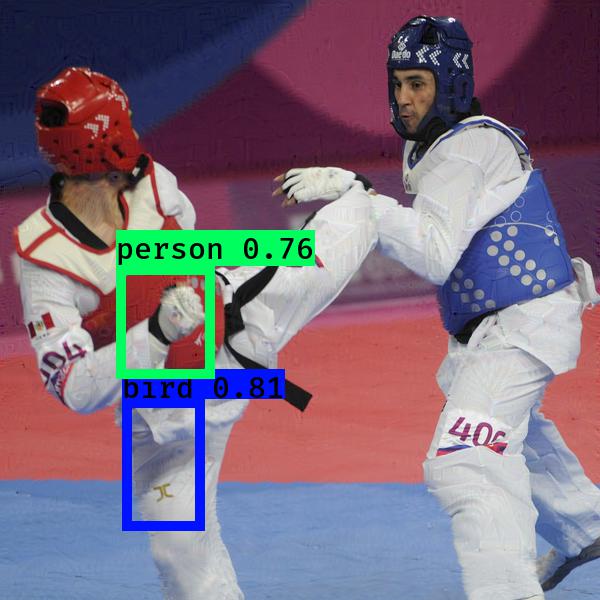} & \includegraphics[scale=0.105]{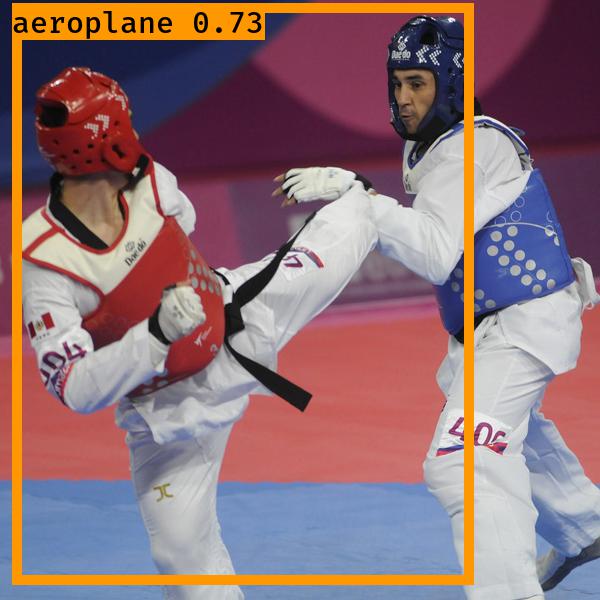} \\
	\end{tabular}
	\vspace{4pt}
	\captionof{figure}{Four visual examples of the untargeted attacks by different algorithms.}
	\label{tab:untargeted-visualization}
	\vspace{-24pt}
\end{table}
Figure~\ref{tab:untargeted-visualization} provides a visualization of four benign images (left most column) and their four adversarial examples generated by TOG, UEA, RAP, and DAG. Four attack algorithms fool the same victim detector FRCNN to misdetect on the same query image in different ways. TOG deceives the victim detector to return false objects on the 1st, 3rd and 4th examples with no correct objects detected. For the 2nd example with two cats, TOG succeeds by fooling the victim to detect no object at all. This shows that different images may respond to the same attack differently, such as missing cats by TOG in the 2nd example compared with fabricating fake objects in the other examples. Similarly, UEA misses both the person and the dog for the 1st example, detects one cat correctly and misses the other cat on the 2nd example, misses both person and car for the 3rd example, and misdetect all objects on the 4th example. RAP and DAG fail the detection on all four examples differently. 

Table~\ref{tab:untargeted-evaluation} provides the quantitative measurements on all victim detectors under the four attack algorithms. 
\begin{table}[t]\renewcommand{\arraystretch}{0.8}
\centerfloat\scriptsize
\setlength\tabcolsep{3.5pt}
\begin{tabular}{@{}ccccccccccc@{}}
	\toprule
	\multirow{2}{*}[-2pt]{\textbf{\begin{tabular}[c]{@{}c@{}}Dataset\end{tabular}}} & \multirow{2}{*}[-2pt]{\textbf{\begin{tabular}[c]{@{}c@{}}Random\\ Attack\end{tabular}}} & \multirow{2}{*}[-2pt]{\textbf{\begin{tabular}[c]{@{}c@{}}Victim\\ Detector\end{tabular}}} & \multicolumn{2}{c}{\textbf{mAP (\%)}} & \multicolumn{2}{c}{\textbf{Time Cost (s)}} & \multicolumn{4}{c}{\textbf{Distortion Cost}} \\ \cmidrule(l){4-5} \cmidrule(l){6-7} \cmidrule(l){8-11} 
	& &  & {Benign} & {Adv.} & {Benign} & {Adv.} & {$L_\infty$} & {$L_2$} & {$L_0$} & {SSIM} \\ \midrule
	VOC & TOG & YOLOv3-D & 83.43 & 0.56 & 0.03 & 0.98 & 0.031 & 0.083 & 0.984 & 0.875 \\
	VOC & TOG & YOLOv3-M & 71.84 & 0.43 & 0.02 & 0.59 & 0.031 & 0.083 & 0.978 & 0.876 \\
	VOC & TOG & SSD300 & 76.11 & 0.86 & 0.02 & 0.39 & 0.031 & 0.120 & 0.975 & 0.879 \\
	VOC & TOG & SSD512 & 79.83 & 0.74 & 0.03 & 0.69 & 0.031 & 0.070 & 0.974 & 0.869 \\
	VOC & TOG & FRCNN & 67.37 & 2.64 & 0.14 & 1.68 & 0.031 & 0.058 & 0.976 & 0.862\\ \midrule
	VOC & UEA & FRCNN & 67.37 & 18.07 & 0.14 & 0.17 & 0.343 & 0.191 & 0.959 & 0.652 \\
	VOC & RAP & FRCNN & 67.37 & 4.78 & 0.14 & 4.04 & 0.082 & 0.010 & 0.531 & 0.994 \\
	VOC & DAG & FRCNN & 67.37 & 3.56 & 0.14 & 7.99 & 0.024 & 0.002 & 0.493 & 0.999\\ \midrule
	COCO & TOG & YOLOv3-D & 54.16 & 3.52 & 0.03 & 1.02 & 0.031 & 0.083 & 0.986 & 0.872 \\ \bottomrule
\end{tabular}
\vspace{4pt}
\caption{Untargeted attacks on different datasets and victim detectors.}
\vspace{-30pt}
\label{tab:untargeted-evaluation}
\end{table}
The first metric is the mAP in percentage, including benign mAP with no attacks and adversarial mAP given adversarial examples. The second metric measures the detection time on benign inputs and attack total cost (both generation and detection). The third metric is the distortion cost measured in $L_\infty$, $L_2$, $L_0$ distances, and SSIM. $L_2$ and $L_0$ costs reported here are normalized by the number of pixels and the $L_2$ cost has a magnitude of $10^{-3}$. Note that UEA, RAP, and DAG can only attack FRCNN, and hence we do not evaluate them on YOLOv3, SSD300 and SSD512. We make two observations from Table~\ref{tab:untargeted-evaluation}. {First}, all attacks successfully bring down the mAP of the victim. Considering the TOG attack, the benign mAP of any victim detector is drastically reduced to less than $3.52\%$ with four victims having a close to zero adversarial mAP. This indicates that the victims fail miserably with no detection capability. {Second}, we compare four different attacks on FRCNN, which has a benign mAP of $67.37\%$. TOG is the most powerful attack with the lowest adversarial mAP of $2.64\%$, followed by DAG ($3.56\%$), RAP ($4.78\%$), and UEA ($18.07\%$). By default, UEA generates adversarial examples with a fixed resolution of $300\times300$. When attacking FRCNN taking inputs with resolution of $600\times600$, resizing and interpolation are required. Hence, the effectiveness of UEA is hindered. In comparison, TOG, RAP and DAG are much more adaptive, and capable of generating adversarial examples that fit the input resolution, as they do not rely on additional networks.

Apart from attack effectiveness, attack costs are equally important. UEA has the lowest time cost with only $0.17$s attack total time because the generation of adversarial examples does not use the victim model but the GAN, which can have much lower complexity. TOG has a reasonable range of attack total time but RAP and DAG have prohibitively high time cost ($4.04$s and $7.99$s). This can be explained by the number of iterations required to succeed the attack in RAP and DAG. TOG needs $10$ iteration while RAP and DAG need to run more than 30 rounds. Interestingly, spending more iterations allows RAP and DAG to have a much lower distortion cost and exceptionally high SSIM measures of $0.994$ and $0.999$ respectively. TOG also has a high imperceptibility with SSIM higher than $0.862$, while adversarial perturbation generated by UEA is significantly more perceptible, having a low SSIM of $0.652$. Furthermore, RAP and DAG have a low $L_0$ cost, which implies their perturbations are more localized. In comparison, both TOG and UEA have the $L_0$ cost close to $1.000$, indicating that most pixels are modified by the adversarial perturbation.
	
\vspace{-8pt}\subsection{Targeted Specificity Attacks}
\begin{figure*}[t]
	\vspace{-8pt}
	\centering
	\begin{subfigure}[t]{0.33\textwidth}
		\centering
		\includegraphics[width=\textwidth]{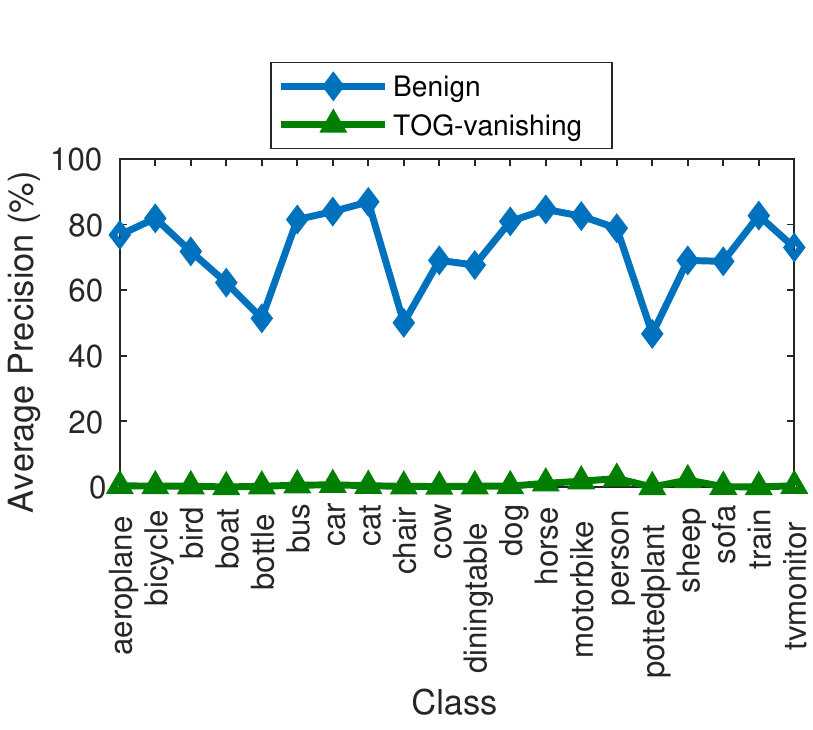}
		\caption{TOG-vanishing}
	\end{subfigure}%
	~ 
	\begin{subfigure}[t]{0.33\textwidth}
		\centering
		\includegraphics[width=\textwidth]{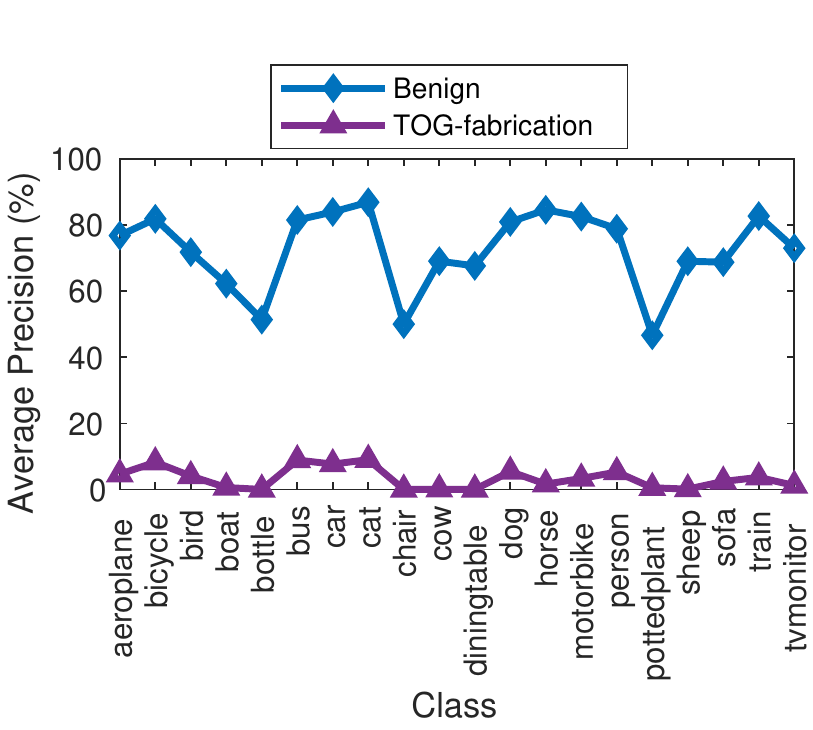}
		\caption{TOG-fabrication}
	\end{subfigure}%
	~ 
	\begin{subfigure}[t]{0.33\textwidth} 
		\centering
		\includegraphics[width=\textwidth]{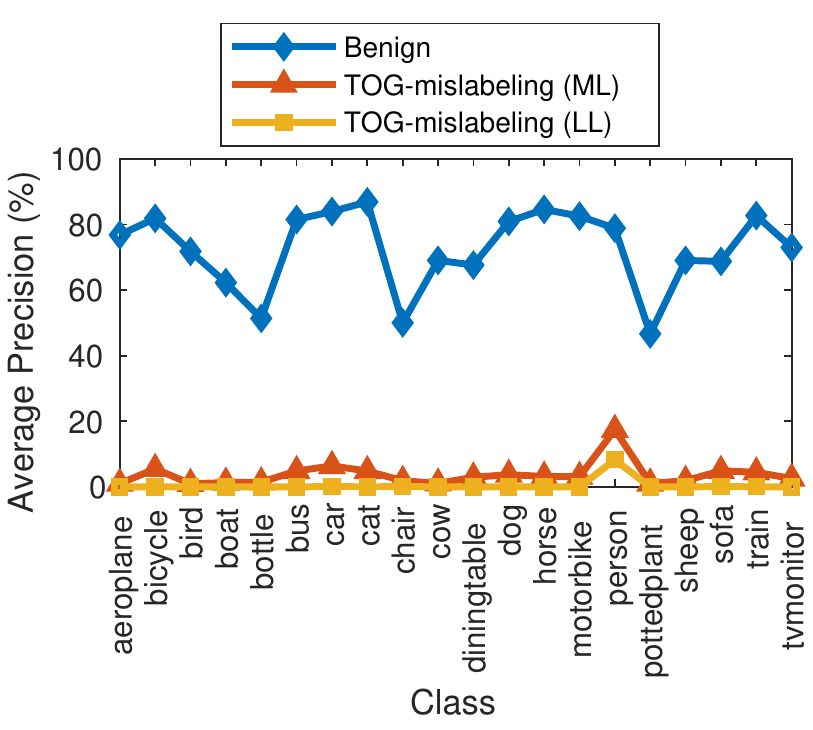}
		\caption{TOG-mislabeling}
	\end{subfigure}%
	\vspace{-6pt}
	\caption{{\small The AP of each class under TOG targeted attacks on YOLOv3-M}}
	\label{fig:exp-voc-yolov3-m}
	\vspace{-12pt}
\end{figure*}

\vspace{-4pt}
We evaluate the three targeted specificity attacks using TOG. For targeted mislabeling attacks, without loss of generality, we choose two representative attack targets: the most-likely (ML) and the least-likely (LL), which correspond to the incorrect class label of an object detected on benign example with the highest and the lowest prediction confidence respectively~\cite{chow2019denoising}. The TOG-mislabeling allows objects of any class to be attacked. Figure~\ref{fig:exp-voc-yolov3-m} shows the benign and adversarial AP of each class on YOLOv3-M. All targeted attacks by TOG drastically reduce the average precision of \emph{every} class supported by the victim to almost zero, showing the severity of the targeted attacks. We provide more experimental measurements on all $24$ cases (four attacks on six detectors) in Appendix B.

Recall Figure~\ref{tab:untargeted-visualization}, each of the four input images responds to the same untargeted random attack differently. Figure~\ref{tab:targeted-visualization} provides a visualization of the same set of images attacked by TOG with different targeted specificity effects. This qualitatively validates that all targeted attacks in TOG are goal-driven, which can be more detrimental to victim detector. For example, with TOG-vanishing attack ($2$nd column), all four adversarial examples fool the victim detector FRCNN to misdetect with no object recognized. For TOG-mislabeling attacks, the person and the dog on the $1$st row are purposefully mislabeled as the dog and the cat respectively in the ML case and both aeroplanes in the LL case. In comparison with Figure~\ref{tab:untargeted-visualization}, UEA, RAP, DAG and general TOG are untargeted: each of the four input images responds to attacks under the same attack algorithm (be it TOG, UEA, RAP and DAG) quite differently, showing random ways to fool a victim detector. We provide more experimental analysis on each targeted attack in Appendix B.
\begin{table}[t]\renewcommand{\arraystretch}{0.8}
\centering\scriptsize
\begin{tabular}{ccccc}
	{\begin{tabular}[c]{@{}c@{}}No Attack\\\;\end{tabular}} & {\begin{tabular}[c]{@{}c@{}}TOG-vanishing\\\;\end{tabular}} & {\begin{tabular}[c]{@{}c@{}}TOG-fabrication\\\;\end{tabular}} & {\begin{tabular}[c]{@{}c@{}}TOG-mislabeling\\(ML)\end{tabular}} & {\begin{tabular}[c]{@{}c@{}}TOG-mislabeling\\(LL)\end{tabular}} \\ 
	\includegraphics[scale=0.1]{figures/untargeted_examples/example1_benign.jpg} & \includegraphics[scale=0.1]{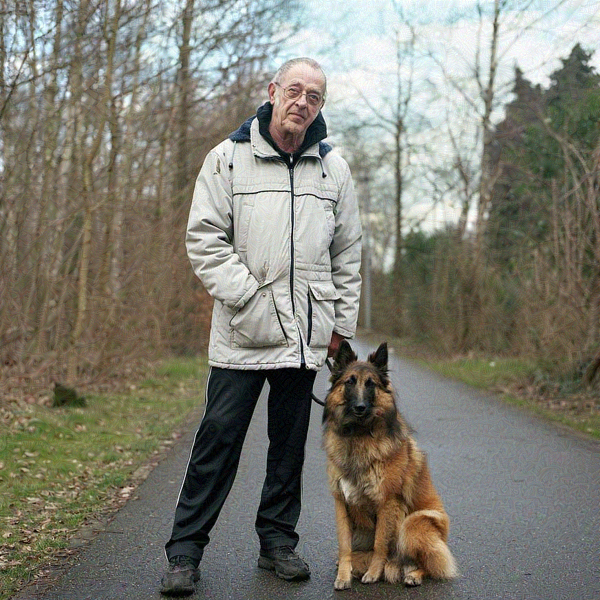} & \includegraphics[scale=0.1]{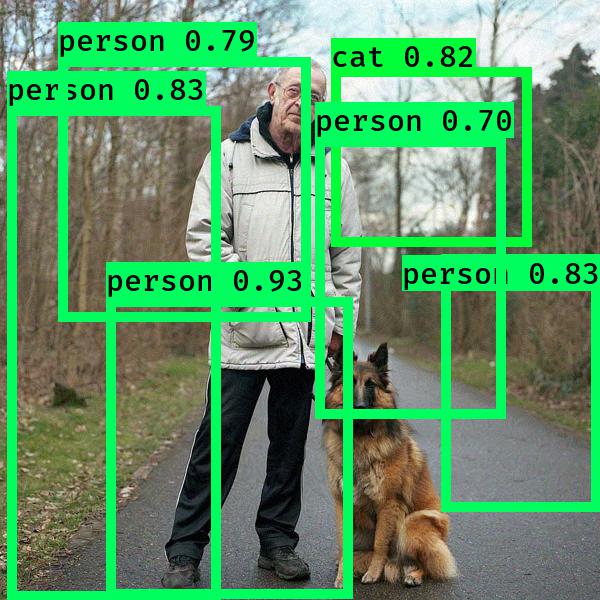} & \includegraphics[scale=0.1]{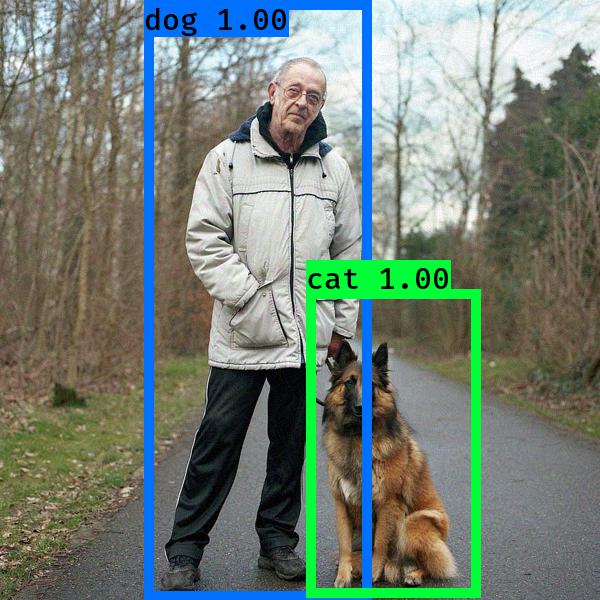} & \includegraphics[scale=0.1]{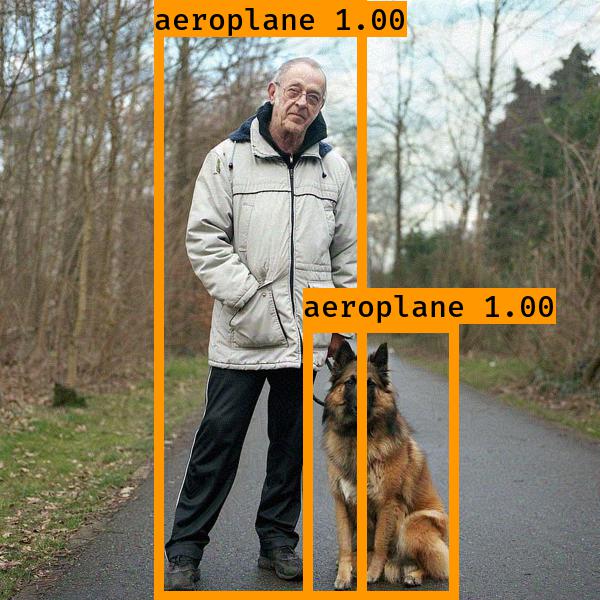} \\
	\includegraphics[scale=0.1]{figures/untargeted_examples/example2_benign.jpg} & \includegraphics[scale=0.1]{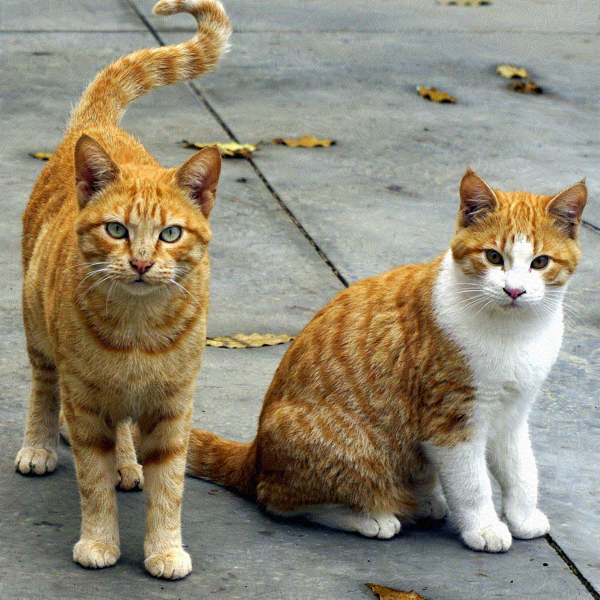} & \includegraphics[scale=0.1]{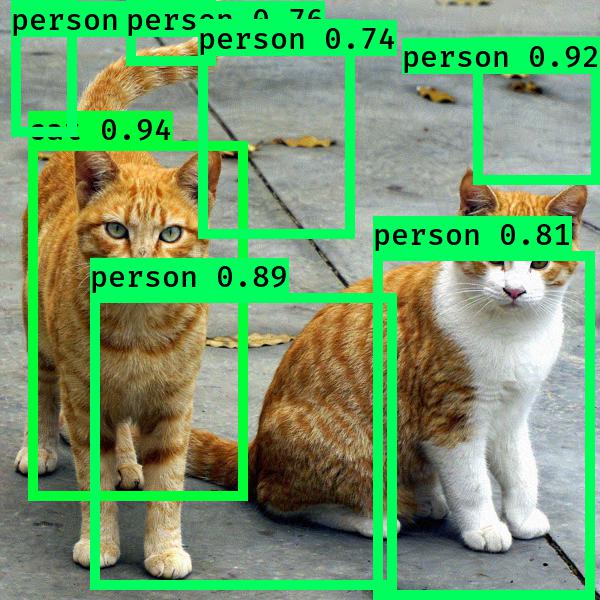} & \includegraphics[scale=0.1]{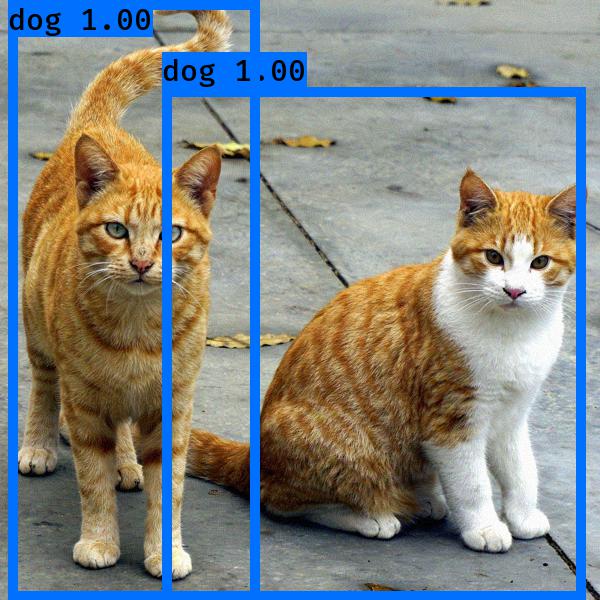} & \includegraphics[scale=0.1]{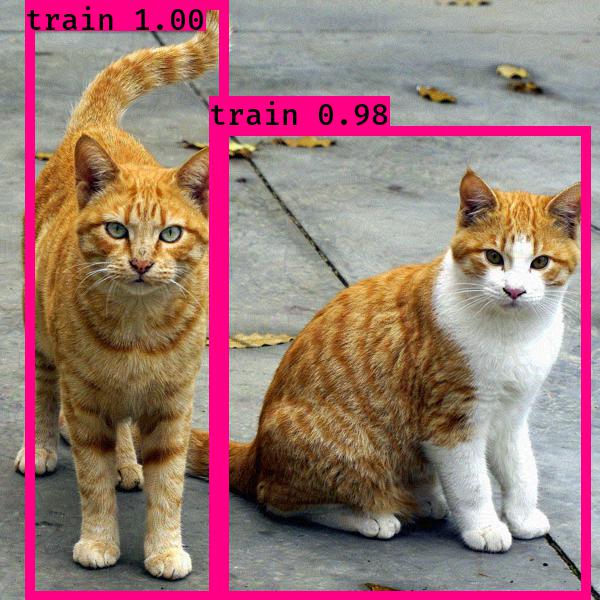} \\
	\includegraphics[scale=0.1]{figures/untargeted_examples/example3_benign.jpg} & \includegraphics[scale=0.1]{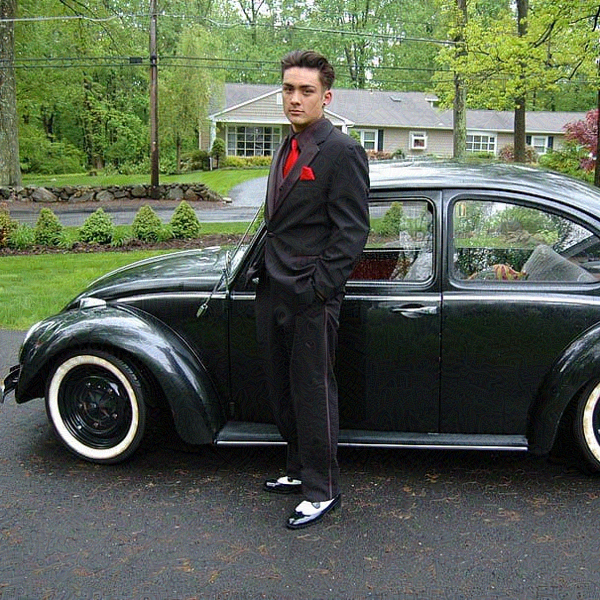} & \includegraphics[scale=0.1]{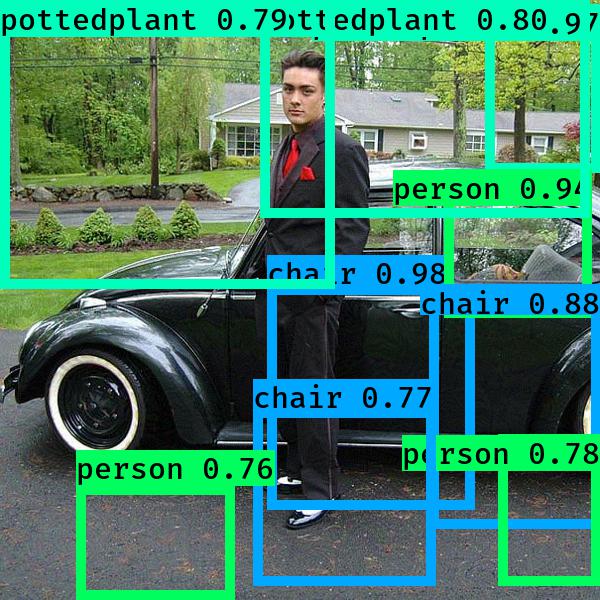} & \includegraphics[scale=0.1]{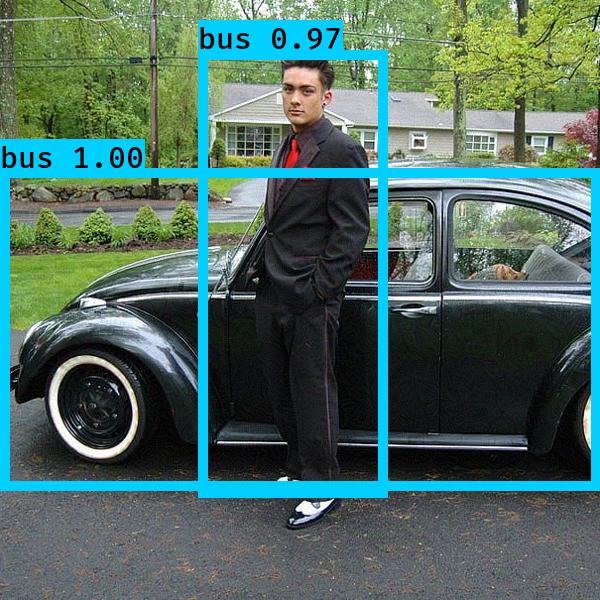} & \includegraphics[scale=0.1]{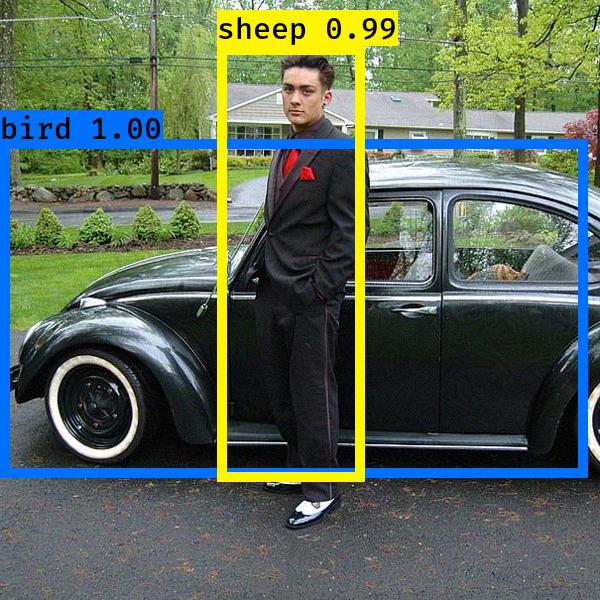} \\
	\includegraphics[scale=0.1]{figures/untargeted_examples/example4_benign.jpg} & \includegraphics[scale=0.1]{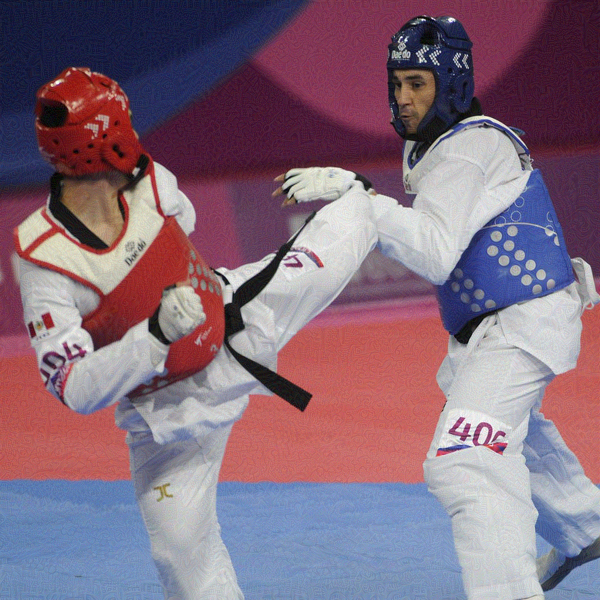} & \includegraphics[scale=0.1]{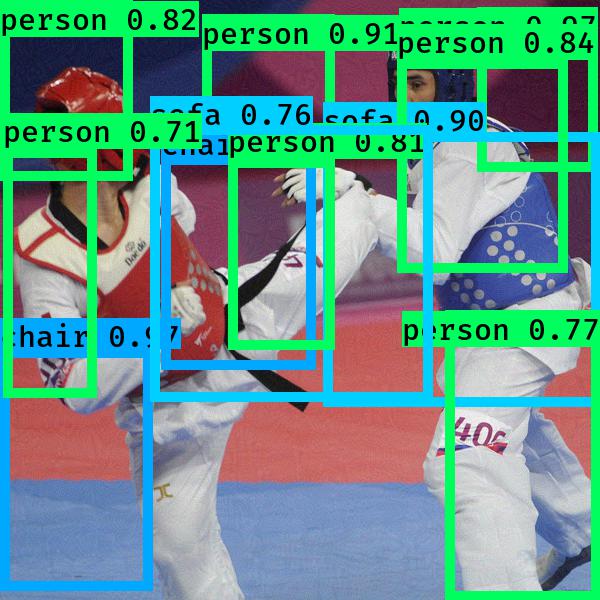} & \includegraphics[scale=0.1]{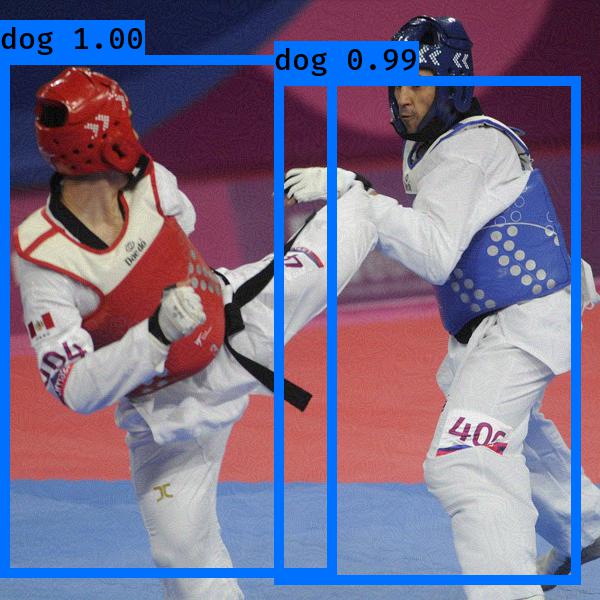} & \includegraphics[scale=0.1]{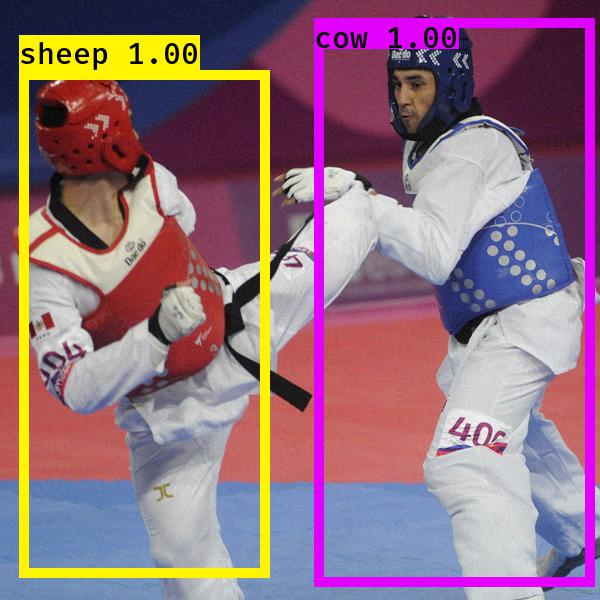} \\
\end{tabular}
\vspace{4pt}
\captionof{figure}{Four visual examples of different targeted specificity attacks by TOG.}
\label{tab:targeted-visualization}
\vspace{-24pt}
\end{table}

\vspace{-8pt}\subsection{Transferability of Attacks}\vspace{-4pt}\label{sec:exp-transferability}
We conduct quantitative analysis on the transferability of all four untargeted attacks: TOG, UEA, RAP and DAG. Table~\ref{tab:cross-backbone-algorithm-transferability} reports the results for the cross-model transferability, measured in adversarial mAP.
\begin{table}[t]\renewcommand{\arraystretch}{0.8}
\centering\scriptsize
\setlength\tabcolsep{3pt}
\begin{tabular}{@{}ccccccc@{}}
	\toprule
	\multirow{2}{*}[-2pt]{\textbf{\begin{tabular}[c]{@{}c@{}}Transfer\\Attack\end{tabular}}} & \multirow{2}{*}[-2pt]{\textbf{\begin{tabular}[c]{@{}c@{}}Source\\Model\end{tabular}}} & \multicolumn{5}{c}{\textbf{Target Model}} \\ \cmidrule(l){3-7} 
	\multicolumn{1}{c}{} &  & YOLOv3-D & YOLOv3-M & SSD300 & SSD512 & FRCNN \\ \midrule
	\multicolumn{2}{c}{Benign (No Attack)} & 83.43 & 71.84 & 76.11 & 79.83 & 67.37 \\ \midrule
	TOG & YOLOv3-D & \textbf{0.56} & 60.13 & 72.70 & 73.86 & 55.57 \\
	TOG & YOLOv3-M & 74.62 & \textbf{0.43} & 73.27 & 75.27 & 59.1  \\
	TOG & SSD300 & 56.87 & 42.85& \textbf{0.86} & 38.79 & 50.36  \\
	TOG & SSD512 & 56.21 & 46.00 & 58.00 & \textbf{0.74} & 35.98 \\
	TOG & FRCNN & 79.47 & 68.60 & 75.80 & 78.09 & \textbf{2.64} \\ \midrule
	UEA & FRCNN & 51.92 & 31.88 & 47.08 & 47.66 & \textbf{18.07} \\ 
	RAP & FRCNN & 81.80 & 69.45 & 75.77 & 76.84 & \textbf{4.78} \\ 
	DAG & FRCNN & 81.21 & 70.37 & 75.15 & 78.38 & \textbf{3.56} \\ \bottomrule
\end{tabular}
\vspace{4pt}
\caption{Cross-model transferability.}
\vspace{-12pt}
\label{tab:cross-backbone-algorithm-transferability}
\end{table}
Using the same model to craft adversarial examples always achieves the highest transferability, as indicated in boldface. We first consider the adversarial examples generated on different source models and measure their transferability to different target models using TOG (the $2$nd-$6$th rows). {First}, we observe that having the same backbone architecture does not necessarily lead to high transferability. FRCNN, SSD300 and SSD512 all use VGG16 as the backbone network. Yet, the adversarial examples generated on FRCNN have very low transferability to SSD300 and SSD512, reducing their mAP from $76.11\%$ to $75.80\%$ and from $79.83\%$ to $78.09\%$ respectively. {Second}, the adversarial examples generated on SSD have relatively higher transferability compared to other source models. For instance, adversarial examples from SSD300 and SSD512 can reduce the mAP of YOLOv3-D from $83.43\%$ to $56.87\%$ and $56.21\%$, much better than YOLOv3-M and FRCNN that only reduction to $74.62\%$ and $79.47\%$ are recorded. {Finally}, considering the transferability of different attack algorithms with the same source model FRCNN (the last four rows), we find that adversarial examples by UEA exhibit a higher transferability consistently. This can be attributed to its high distortion cost incurred to perturb each adversarial example (recall Table~\ref{tab:untargeted-evaluation}). 

Table~\ref{tab:frcnn-res} and Table~\ref{tab:yolov3-res} report the cross-resolution transferability on FRCNN and YOLOv3 respectively. 
\begin{table}[t]\renewcommand{\arraystretch}{0.8}
\centerfloat
\begin{subtable}{\textwidth}
	\centerfloat
	\setlength\tabcolsep{3pt}\scriptsize
	\begin{tabular}{@{}cccccccc@{}}
		\toprule
		\multirow{2}{*}[-2pt]{\textbf{\begin{tabular}[c]{@{}c@{}}Transfer\\Attack\end{tabular}}} &  \multirow{2}{*}[-2pt]{\textbf{\begin{tabular}[c]{@{}c@{}}Source\\Resolution\end{tabular}}} & \multicolumn{6}{c}{\textbf{Target Resolution}} \\ \cmidrule(l){3-8} 
		\multicolumn{1}{c}{} & & 300x300 & 400x400 & 500x500 & 600x600 & 700x700 & 800x800 \\ \midrule
		\multicolumn{2}{c}{Benign (No Attack)} & 65.33 & 67.85 & 68.00 & 67.37 & 67.91  & 67.76  \\ \midrule
		TOG  & 600x600 & 50.15 & 29.50 & 15.07 & \textbf{2.64} & 6.84  & 3.86  \\
		UEA  & 300x300 & \textbf{3.86} & 11.88 & 18.61 & 18.07 & 16.32 & 17.34 \\ 
		RAP  & 600x600 & 58.45 & 54.32 & 56.96 & \textbf{4.78} & 53.21 & 50.12 \\ 
		DAG  & 600x600 & 62.89 & 59.82 & 46.58 & \textbf{2.84}  & 30.96 & 13.75 \\ \bottomrule
	\end{tabular}
	\vspace{-2pt}
	\caption{\scriptsize FRCNN}
	\vspace{-6pt}
	\label{tab:frcnn-res}
\end{subtable}
\begin{subtable}{\textwidth}
	\centerfloat
	\setlength\tabcolsep{3pt}\scriptsize
	\begin{tabular}{@{}cccccccc@{}}
		\toprule
		\multirow{2}{*}[-2pt]{\textbf{\begin{tabular}[c]{@{}c@{}}Model\end{tabular}}} &\multirow{2}{*}[-2pt]{\textbf{\begin{tabular}[c]{@{}c@{}}Transfer\\Attack\end{tabular}}} &  \multirow{2}{*}[-2pt]{\textbf{\begin{tabular}[c]{@{}c@{}}Source\\Resolution\end{tabular}}} & \multicolumn{5}{c}{\textbf{Target Resolution}} \\ \cmidrule(l){4-8} 
		\multicolumn{1}{c}{} &  & & 352x352 & 384x384 & 416x416 & 448x448 & 480x480 \\ \midrule
		\multirow{2}{*}{YOLOv3-D}& \multicolumn{2}{c}{Benign (No Attack)} & 82.71 & 83.25 & 83.43  & 83.63  & 83.65 \\ 
		& TOG & 416x416 & 25.26 & 14.93 & \textbf{0.56}  & 11.02  & 12.16 \\ \midrule
		\multirow{2}{*}{YOLOv3-M}& \multicolumn{2}{c}{Benign (No Attack)} & 69.98 & 71.13 & 71.84 & 73.10 & 72.72 \\ 
		& TOG & 416x416 & 33.41 & 20.61 & \textbf{0.43}  & 15.62  & 19.16 \\ \bottomrule
	\end{tabular}
	\vspace{-2pt}
	\caption{\scriptsize YOLOv3}
	\label{tab:yolov3-res}
\end{subtable}
\vspace{-6pt}
\caption{Cross-resolution transferability.}
\vspace{-24pt}
\label{tab:cross-resolution-transferability}
\end{table}
Note that only TOG can directly attack YOLOv3 (one-phase detectors), and SSD does not support variable input resolutions. We use nearest neighbor interpolation during resizing as we find empirically that it can better preserve the malicious pattern. For victim detector FRCNN, we observe that TOG and UEA have higher cross-resolution transferability than RAP and DAG. The same observation can be made in both YOLOv3 detectors. For instance, TOG can still effectively reduce the mAP from more than $82\%$ to less than $26\%$ in all target resolutions evaluated on YOLOv3-D. This is because adversarial examples generated by TOG and UEA have a higher robustness under resizing and interpolation to fit the target resolution. Also, upsizing to a higher target resolution is always better than downsizing, causing a higher mAP drop in the target victim model, which can be explained by the fact that downsizing loses the fine details of malicious perturbation.

Table~\ref{tab:online-attack-showcase} provides a visualization to illustrate the transferability of four TOG targeted attacks by generating adversarial examples on SSD300 and evaluating their cross-model transferability to the other three detectors: SSD512, YOLOv3-D, and YOLOv3-M.
\begin{table}[t]\renewcommand{\arraystretch}{0.5}
\centering\scriptsize
\begin{tabular}{@{}cccccc@{}}
	\toprule
	& \multicolumn{5}{c}{\textbf{Detection results under four TOG targeted attacks}} \\ \midrule
	& \begin{tabular}[c]{@{}c@{}}Benign\\ (No Attack)\end{tabular} & TOG-vanishing & TOG-fabrication & \begin{tabular}[c]{@{}c@{}}TOG-mislabeling\\ (ML)\end{tabular} & \begin{tabular}[c]{@{}c@{}}TOG-mislabeling\\ (LL)\end{tabular} \\ \midrule
	\rot{\;\;\;\;\;\;\;\;SSD300} & \includegraphics[scale=0.20]{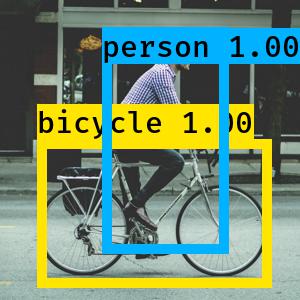} & \includegraphics[scale=0.20]{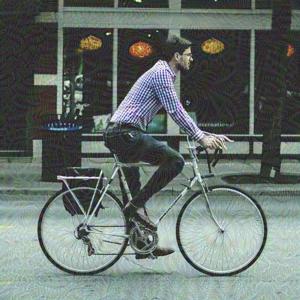} & \includegraphics[scale=0.20]{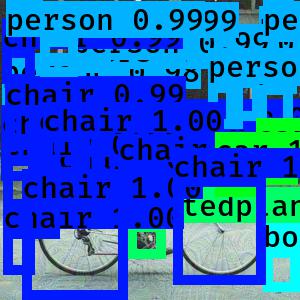} & \includegraphics[scale=0.20]{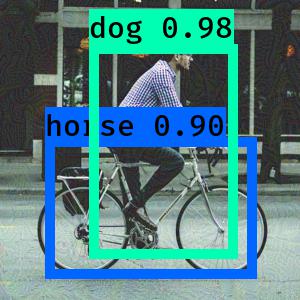} & \includegraphics[scale=0.20]{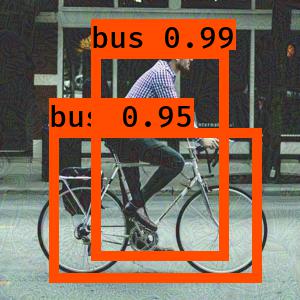} \\ 
	& & &  & \\\toprule
	& \multicolumn{5}{c}{\textbf{Detection results \emph{transferred} from SSD300 to other victim detectors}} \\ \midrule
	& \begin{tabular}[c]{@{}c@{}}Benign\\ (No Attack)\end{tabular} & TOG-vanishing & TOG-fabrication & \begin{tabular}[c]{@{}c@{}}TOG-mislabeling\\ (ML)\end{tabular} & \begin{tabular}[c]{@{}c@{}}TOG-mislabeling\\ (LL)\end{tabular} \\ \midrule
	\rot{\;\;\;\;\;\;\;SSD512} & \includegraphics[scale=0.20]{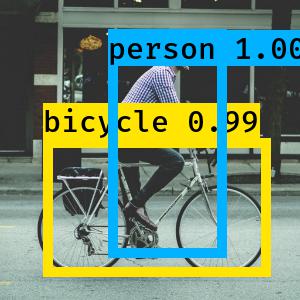} & \includegraphics[scale=0.20]{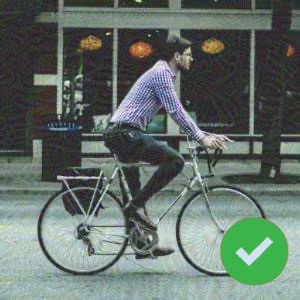} & \includegraphics[scale=0.20]{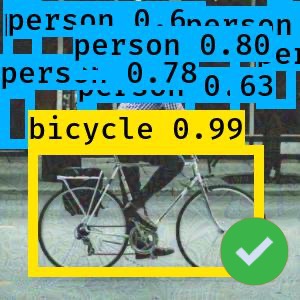} & \includegraphics[scale=0.20]{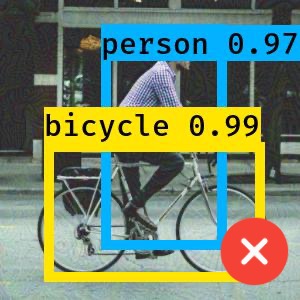} & \includegraphics[scale=0.20]{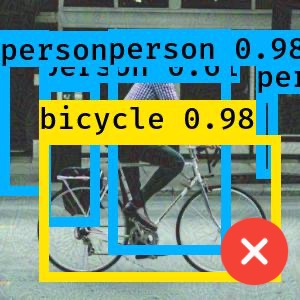} \\
	\rot{\;\;\;\;\;YOLOv3-D} & \includegraphics[scale=0.20]{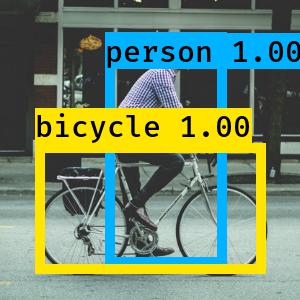} & \includegraphics[scale=0.20]{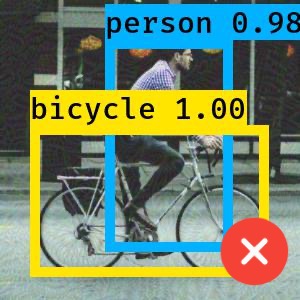} & \includegraphics[scale=0.20]{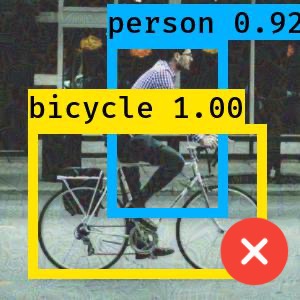} & \includegraphics[scale=0.20]{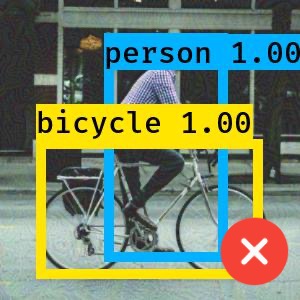} & \includegraphics[scale=0.20]{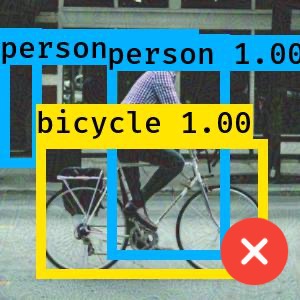} \\
	\rot{\;\;\;\;\;YOLOv3-M} & \includegraphics[scale=0.20]{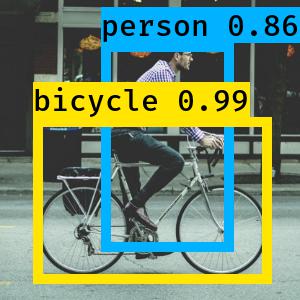} & \includegraphics[scale=0.20]{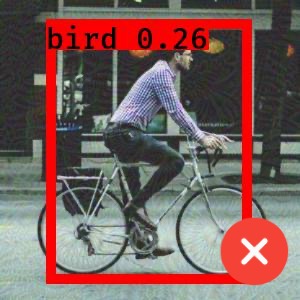} & \includegraphics[scale=0.20]{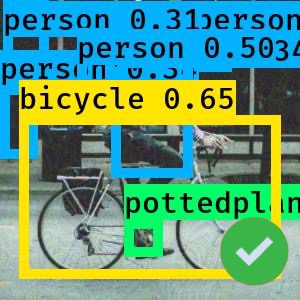} & \includegraphics[scale=0.20]{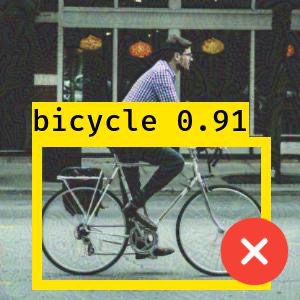} & \includegraphics[scale=0.20]{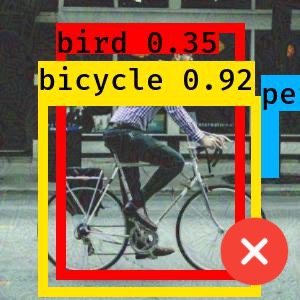} \\ \bottomrule
\end{tabular}
\vspace{4pt}
\caption{Transferring targeted attacks on SSD300 to three other detectors.}
\vspace{-16pt}
\label{tab:online-attack-showcase}
\end{table}
Consider the SSD300 row, the detector can correctly identify the person and the bicycle on the benign input ($1$st column). The targeted attacks by TOG successfully fool the victim to misdetect with designated attack specificity effects: the two objects are missed in TOG-vanishing, false objects are detected in TOG-fabrication, and the person and the bicycle are mislabeled as a dog and a horse in the ML case of TOG-mislabeling and both buses in the LL case. We analyze the transferability by observing the other three rows. Given that all three detectors can successfully identify the two objects on the benign image, we find different degrees of adversarial transferability. For instance, TOG-vanishing and TOG-fabrication can be successfully transferred to SSD512, which has the same backbone (i.e., VGG16) and detection algorithm as the source detector SSD300. TOG-fabrication can also be transferred to YOLOv3-M with the same effect. However, even some adversarial examples may fool other detectors (e.g., TOG-vanishing to YOLOv3-M), they fail in transferring attacks with the same effect. Note that with adversarial transferability, the attacks are black-box, generated and launched without any prior knowledge of the three victim detectors. We provide more discussion in Appendix C.

\vspace{-8pt}\subsection{Model Applicability and Physical Attacks}\vspace{-4pt}
We provide a comparison of seven representative attack algorithms, including two physical attacks, to deep object detectors in Table~\ref{tab:attack-categorization}.
\begin{table}[t]\renewcommand{\arraystretch}{0.8}
\centering\scriptsize
\setlength\tabcolsep{3pt}
\begin{tabular}{@{}lccccccc@{}}
	\toprule
	& \multicolumn{4}{c}{\textbf{Attack Effect}} & \multicolumn{3}{c}{\textbf{Model-applicability}} \\ \cmidrule(l){2-5}  \cmidrule(l){6-8} 
	& \multirow{2}{*}[-2pt]{\scriptsize{Random}} & \multirow{2}{*}[-2pt]{\scriptsize{\begin{tabular}[c]{@{}c@{}}Object-\\ vanishing\end{tabular}}} & \multirow{2}{*}[-2pt]{\scriptsize{\begin{tabular}[c]{@{}c@{}}Object-\\ fabrication\end{tabular}}} & \multirow{2}{*}[-2pt]{\scriptsize{\begin{tabular}[c]{@{}c@{}}Object-\\ mislabeling\end{tabular}}} & \scriptsize{Two-phase} & \multicolumn{2}{c}{\scriptsize{One-phase}}  \\ \cmidrule(l){6-6} \cmidrule(l){7-8} 
	&  &  &  & & \scriptsize{FRCNN} &  \scriptsize{YOLO} & \scriptsize{SSD} \\ \midrule
	TOG~\cite{chow2020tog} & \cmark & \cmark & \cmark & \cmark & \cmark & \cmark & \cmark \\		
	UEA~\cite{wei2018transferable} & \cmark & \xmark & \xmark & \xmark & \cmark & \xmark &  \xmark \\
	
	RAP~\cite{li2018robust} & \cmark & \xmark & \xmark & \xmark  & \cmark & \xmark & \xmark \\
	DAG~\cite{xie2017adversarial} & \cmark & \xmark & \xmark & \xmark  & \cmark & \xmark & \xmark \\ \midrule
	DPATCH~\cite{liu2018dpatch} & \xmark & \xmark & \cmark & \xmark & \cmark & \cmark & \cmark \\
	Extended-RP$_2$~\cite{eykholt2018physical} & \xmark & \cmark & \cmark & \xmark & \cmark & \cmark & \cmark \\	
	Thys's Patch~\cite{thys2019fooling} & \xmark & \cmark & \xmark & \xmark & \cmark & \cmark & \cmark \\	\bottomrule	
\end{tabular}
\vspace{4pt}
\caption{Characteristics of seven representative attacks.}
\vspace{-22pt}
\label{tab:attack-categorization}
\end{table}

TOG~\cite{chow2020tog}, UEA~\cite{wei2018transferable}, RAP~\cite{li2018robust} and DAG~\cite{xie2017adversarial} are the representative digital attacks against a victim detector by perturbing pixel values of a benign image while maximizing one or more of the three loss functions: objectness, bounding box, and classification. All four can perform untargeted random attacks, and TOG also provides additional three targeted specificity attacks. For model-applicability, UEA, RAP and DAG by design depend on the RPN structure, and can only be employed to generate adversarial examples against FRCNN (two-phase detectors). TOG is a general attack framework without dependency on any special structure and can be used to fool object detectors from both one-phase (YOLO and SSD families) and two-phase algorithms (e.g., FRCNN).  

In addition to perturbing the entire image, adversarial patches are also proposed in either a digital (DPATCH) or physical (Extended-RP$_2$ and Thys's Patch) form. DPATCH puts a small patch (e.g., $40\times40$) on a benign example, fooling the victim to fabricate objects at random position or the location where the patch is placed. Extended-RP$_2$ and Thys's Patch propose printable adversarial patches. If the adversarial patch is presented physically in the scene captured by the camera, the captured image will become adversarial input, which will fool a victim detector to misdetect. Extended-RP$_2$ supports ``disappearance'' and ``creation'', corresponding to the object-vanishing and object-fabrication effects, while Thys's Patch aims to make the object vanishing from the detector. Similar to TOG, all physical attack and digital patch algorithms can be employed on both two-phase and one-phase detection techniques.

\vspace{-10pt}\section{Conclusion}\vspace{-6pt}\label{sec:conclusion}
We witnessed a growing number of digital or physical adversarial attacks to object detection systems recently~\cite{chow2020tog,eykholt2018physical,li2018robust,liu2018dpatch,thys2019fooling,wei2018transferable,xie2017adversarial}. To gain an in-depth understanding of the security risks of employing object detection intelligence in security-critical applications, in this paper, we develop a principled evaluation framework to analyze vulnerabilities of object detection systems through an adversarial lens, with three original contributions. First, we examine and compare the state-of-the-art attacks through our proposed evaluation framework. Second, to provide broader coverage of security risks in deep object detection systems, we present a family of TOG attack algorithms, capable of attacking both proposal-based two-phase detectors (e.g., FRCNN) and regression-based one-phase techniques (e.g., SSD, YOLOv3), supporting a general form of untargeted random attacks, and three targeted attacks, geared specifically to object detection. Third but not least, we introduce a set of quantitative metrics, including cross-resolution transferability and cross-model transferability w.r.t. algorithms and DNN backbones,  to evaluate the effectiveness and cost of four representative methods of digital attacks, and using model-applicability to compare digital attacks with physical patch attacks. Our evaluation framework can serve as a tool for analyzing adversarial attacks, assessing security risks and adversarial robustness of deep object detectors deployed in real-world applications. 

\vspace{-12pt}\section*{Acknowledgment}\vspace{-8pt}
This research is partially sponsored by NSF CISE SaTC 1564097 and an IBM faculty award. Any opinions, findings, and conclusions or recommendations expressed in this material are those of the author(s) and do not necessarily reflect the views of the National Science Foundation or other funding agencies and companies mentioned above.

\vspace{-8pt}
\bibliographystyle{splncs04}
\bibliography{ref}

\appendix

\section*{Appendix}\vspace{-4pt}
\textbf{A. Background. \/} The VOC 2007+2012 dataset has $16,551$ training images and $4,952$ testing images, while the COCO 2014 dataset has $117,264$ training images and $5,000$ testing images. The configuration and detection performance of the six detectors under no attack are reported in Table~\ref{tab:victim-summary}. All measurements are recorded on NVIDIA RTX 2080 SUPER (8 GB) GPU, Intel i7-9700K (3.60GHz) CPU, and 32 GB RAM on Ubuntu 18.04.
\begin{table}
	\vspace{-8pt}
	\centering\scriptsize
	\setlength\tabcolsep{4pt}
	\begin{tabular}{@{}ccccccc@{}}
		\toprule
		\textbf{Dataset} & \textbf{\begin{tabular}[c]{@{}c@{}}Detector\\ Identifier\end{tabular}} & \textbf{Algorithm} & \textbf{Backbone} & \textbf{\begin{tabular}[c]{@{}c@{}}Input\\ Resolution\end{tabular}} & \textbf{\begin{tabular}[c]{@{}c@{}}Benign\\ mAP(\%)\end{tabular}} & \textbf{\begin{tabular}[c]{@{}c@{}}Detection\\ Time(s)\end{tabular}} \\ \midrule
		\multirow{5}{*}{VOC} & YOLOv3-D & YOLOv3 & Darknet53 & 416x416 & 83.43 & 0.0328\\
		& YOLOv3-M & YOLOv3 & MobileNetV1 & 416x416 & 71.84 & 0.0152  \\
		& SSD300 & SSD & VGG16 & 300x300 & 76.11 & 0.0208 \\
		& SSD512 & SSD & VGG16 & 512x512 & 79.83 & 0.0330 \\
		& FRCNN & Faster R-CNN & VGG16 & 600x600 & 67.37 & 0.1399 \\ \midrule
		COCO & YOLOv3-D & YOLOv3 & Darknet53 & 416x416 & 54.16 & 0.0337 \\ \bottomrule
	\end{tabular}
	\vspace{4pt}
	\caption{A summary of victim detectors under no attack.}
	\vspace{-20pt}
	\label{tab:victim-summary}
\end{table}

\textbf{B. Analysis on Targeted Specificity Attacks. \/} 
Table~\ref{tab:targeted-tog-evaluation} reports the results of four TOG targeted attacks on six victim detectors ($24$ cases). 
\begin{table}[t]
	\centerfloat\scriptsize
	\setlength\tabcolsep{3pt}
	\begin{tabular}{@{}clcccccccc@{}}
		\toprule
		\multirow{2}{*}[-2pt]{\textbf{\begin{tabular}[c]{@{}c@{}}Detector\\(Dataset) \end{tabular}}} & \multicolumn{1}{c}{\multirow{2}{*}[-2pt]{\textbf{Targeted Attack}}} & \multicolumn{2}{c}{\textbf{mAP (\%)}} & \multicolumn{2}{c}{\textbf{Time Cost (s)}} & \multicolumn{4}{c}{\textbf{Distortion Cost}} \\ \cmidrule(l){3-4}  \cmidrule(l){5-6}  \cmidrule(l){7-10} 
		& \multicolumn{1}{c}{} & {Benign} & {Adv.} & {Benign} & {Adv.} & {$L_\infty$} & {$L_2$} & {$L_0$} & {SSIM} \\  \midrule
		\multirow{4}{*}{\begin{tabular}[c]{@{}c@{}}YOLOv3-D\\(VOC)\end{tabular}} & TOG-vanishing & 83.43 & 0.32 & 0.03 & \textbf{0.77} & 0.031 & 0.082 & 0.983 & 0.877 \\
		& TOG-fabrication & 83.43  & \textbf{0.25} & 0.03 & 0.93 & 0.031 & 0.084 & 0.984 & {0.873} \\
		& TOG-mislabeling (ML) & 83.43 & 3.15 & 0.03 & 0.95& 0.031 & 0.080 & 0.972 & \textbf{0.879} \\
		& TOG-mislabeling (LL) & 83.43 & 2.80 & 0.03 & 0.96& 0.031 & 0.081 & 0.972 & \textbf{0.879} \\ \midrule
		\multirow{4}{*}{\begin{tabular}[c]{@{}c@{}}YOLOv3-M\\(VOC)\end{tabular}} & TOG-vanishing & 71.84 & 0.36 & 0.02 & \textbf{0.37}& 0.031 & 0.082 & 0.978 & {0.878} \\
		& TOG-fabrication & 71.84 & \textbf{0.17} & 0.02 & 0.57& 0.031 & 0.084 & 0.976 & {0.873} \\
		& TOG-mislabeling (ML) & 71.84 & 2.67 & 0.02 & 0.56& 0.031 & 0.079 & 0.953 & \textbf{0.882} \\
		& TOG-mislabeling (LL) & 71.84 & 1.60 & 0.02 & 0.56& 0.031 & 0.079 & 0.953 & 0.881 \\ \midrule
		\multirow{4}{*}{\begin{tabular}[c]{@{}c@{}}SSD300\\(VOC)\end{tabular}} & TOG-vanishing & 76.11 & 5.54& 0.02 & \textbf{0.36}& 0.031 & 0.120 & 0.978 & 0.880 \\
		& TOG-fabrication & 76.11 & \textbf{0.57} & 0.02 & 0.37& 0.031 & 0.122 & 0.978 & {0.877} \\ 
		& TOG-mislabeling (ML) & 76.11 & 2.53 & 0.02 & 0.37& 0.030 & 0.110 & 0.945 & \textbf{0.891} \\
		& TOG-mislabeling (LL) & 76.11 & 1.44 & 0.02 & 0.37& 0.030 & 0.111 & 0.945 & 0.889 \\ \midrule
		\multirow{4}{*}{\begin{tabular}[c]{@{}c@{}}SSD512\\(VOC)\end{tabular}} & TOG-vanishing & 79.83 & 6.23 & 0.03 & \textbf{0.62} & 0.031 & 0.071 & 0.975 & 0.868 \\
		& TOG-fabrication & 79.83 & \textbf{0.50} & 0.03 & 0.69& 0.031 & 0.071 & 0.976 & {0.866} \\
		& TOG-mislabeling (ML) & 79.83 & 2.53 & 0.03 & 0.65 & 0.031 & 0.065 & 0.957 & \textbf{0.878} \\
		& TOG-mislabeling (LL) & 79.83 & 1.20 & 0.03 & 0.65 & 0.031 & 0.066 & 0.956 & 0.877 \\ \midrule
		\multirow{4}{*}{\begin{tabular}[c]{@{}c@{}}FRCNN\\(VOC)\end{tabular}} & TOG-vanishing & 67.37 & \textbf{0.14} & 0.14 & \textbf{1.66}& 0.031 & 0.058 & 0.975 & {0.862} \\
		& TOG-fabrication & 67.37 & 1.24 & 0.14 & 1.68& 0.031 & 0.057 & 0.977 & 0.866 \\
		& TOG-mislabeling (ML) & 67.37 & 2.14 & 0.14 & 1.64& 0.030 & 0.054 & 0.935 & \textbf{0.873} \\
		& TOG-mislabeling (LL) & 67.37 & 1.44 & 0.14 & 1.60& 0.030 & 0.054 & 0.935 & 0.872 \\ \midrule
		\multirow{4}{*}{\begin{tabular}[c]{@{}c@{}}YOLOv3-D\\(COCO)\end{tabular}} & TOG-vanishing & 54.16 & \textbf{0.41} & 0.03 & \textbf{0.78} & 0.031 & 0.082 & 0.986 & 0.874 \\
		& TOG-fabrication & 54.16 & 1.46 & 0.03 &\textbf{0.78}& 0.031 & 0.083 & 0.986 & {0.871} \\
		& TOG-mislabeling (ML) & 54.16 & 5.43 & 0.03 & 1.00& 0.031 & 0.080 & 0.968 & \textbf{0.878} \\
		& TOG-mislabeling (LL) & 54.16 & 0.76 & 0.03 & 1.00& 0.031 & 0.080 & 0.968 & 0.877 \\ \bottomrule
	\end{tabular}
	\vspace{4pt}
	\caption{Targeted attacks by TOG on different datasets and victim detectors.}
	\vspace{-20pt}
	\label{tab:targeted-tog-evaluation}
\end{table}
TOG targeted attacks effectively bring down the mAP of all victim detectors, with any attack specificity. For instance, YOLOv3-D on VOC has a high mAP of $83.43\%$ given benign images but, under attacks, it  becomes less than $3.15\%$. Even though the adversarial examples in targeted attacks can fool the victim detectors to misdetect with the targeted specificity effects, such attack sophistication does not drastically incur additional attack time cost and distortion cost, compared with the TOG untargeted attack scenario in Table~\ref{tab:untargeted-evaluation}.

\begin{figure}
	\centering
	\includegraphics[width=0.80\textwidth]{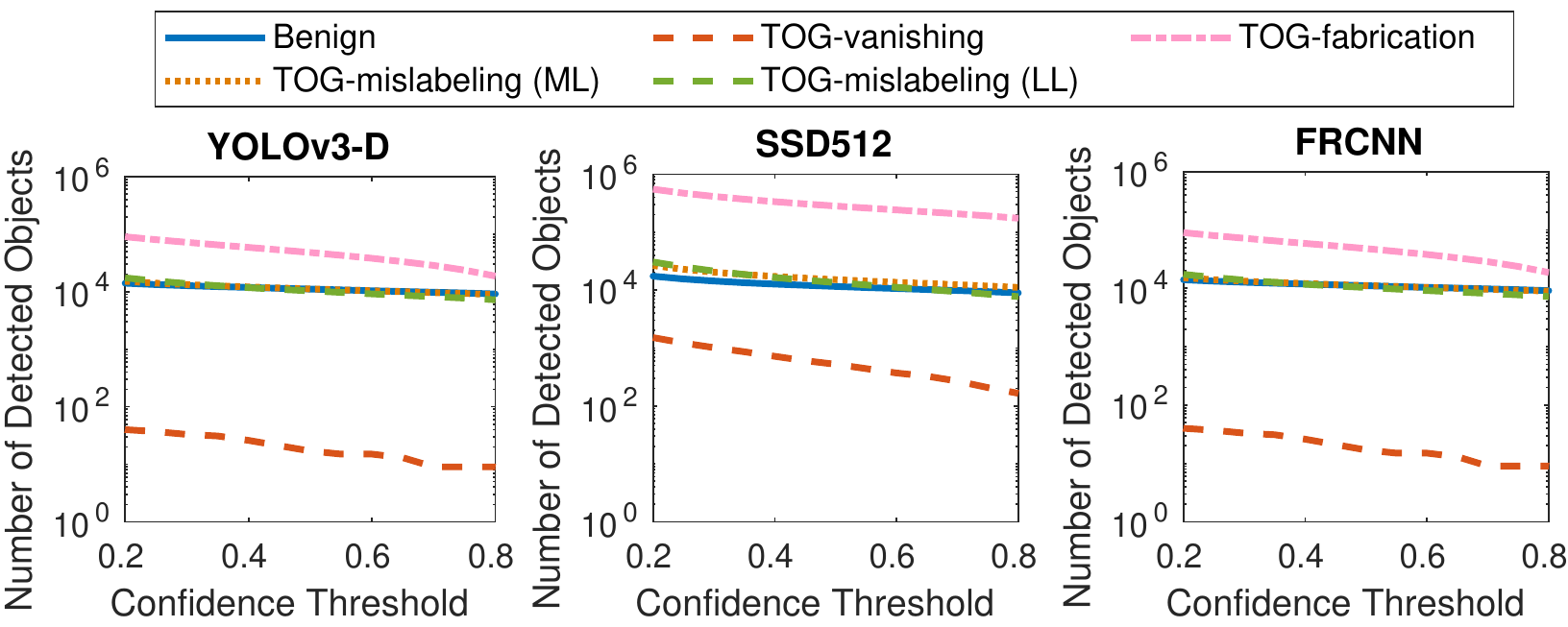}
	\vspace{-6pt}
	\caption{Number of detected objects under no attack and TOG targeted attacks.}
	\vspace{-12pt}
	\label{fig:tog-num-objects}
\end{figure}

Figure~\ref{fig:tog-num-objects} compares the four targeted attacks with respect to the number of object detected by three victim detectors (YOLOv3-D, SSD512 and FRCNN) with different settings of the confidence threshold. The benign case (the blue solid curve) indicates the number of objects detected by the victims under no attacks. Confidence thresholding is used by object detection algorithms as a post-processing step to return only detected objects with high confidence (Section~\ref{sec:overview-od}), and the threshold is a hyperparameter defined by the system owner (e.g,. FRCNN uses $0.70$ by default). We find that all trends are consistent across both detectors: Figure~\ref{fig:tog-num-objects} experimentally confirms that (i) the TOG-vanishing attacks significantly lower the number of detected objects with any setting of confidence threshold, (ii) the number of detected objects is drastically increased in TOG-fabrication attacks, and (iii) the TOG-mislabeling attacks (both ML and LL) have almost the same number of objects detected on benign examples. 

\begin{figure}
\centering
\vspace{-8pt}
\begin{subfigure}{0.45\textwidth}
	\includegraphics[width=\textwidth]{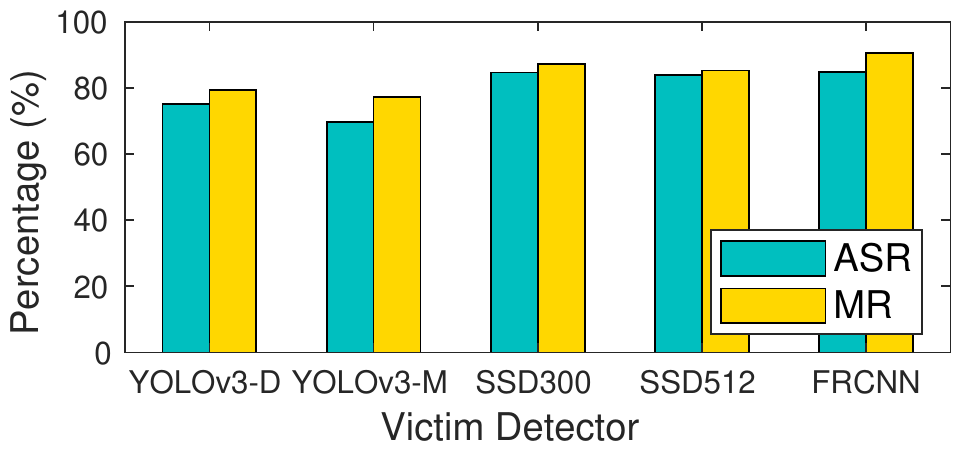}
	\vspace{-12pt}
	\caption{\scriptsize Most-likely (ML) Targets}
	\label{fig:asr_mr_mislabeling-ml}
\end{subfigure}\hspace{2pt}
\begin{subfigure}{0.45\textwidth}
	\includegraphics[width=\textwidth]{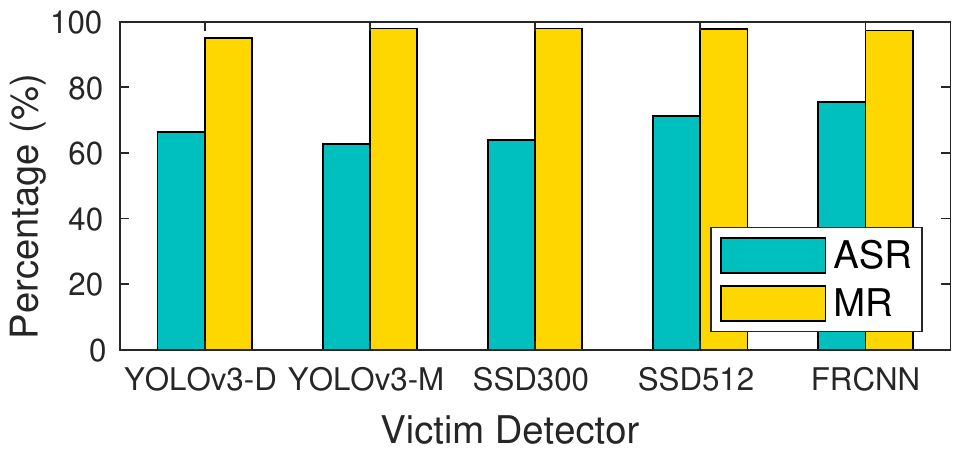}
	\vspace{-12pt}
	\caption{\scriptsize Least-likely (LL) Targets}
	\label{fig:asr_mr_mislabeling-ll}
\end{subfigure}
\vspace{-6pt}
\caption{ASR and MR of TOG-mislabeling attacks.}
\vspace{-8pt}
\label{fig:asr_mr_mislabeling}
\end{figure}

Figure~\ref{fig:asr_mr_mislabeling} further analyzes the two targeted mislabeling attacks of TOG in terms of ASR according to Equation~\ref{eq:asr-mislabeling}. With a similar formulation, we also introduce misdetection rate (MR) to compute the portion of objects that are mislabeled under TOG-mislabeling attacks. Note that MR still requires the detected bounding box to be correct, but the predicted class label of the object can be any class but not the correct one. We observe that a large portion of objects are successfully mislabeled as the maliciously targeted class (ASR), and only small portion is randomly mislabeled instead (MR - ASR), especially for the ML targets (Figure~\ref{fig:asr_mr_mislabeling-ml}). For the LL attack targets (Figure~\ref{fig:asr_mr_mislabeling-ll}), the ASR is less than 80\%, but the misdetection rate (MR) is close to $100\%$ in all five victim detectors, indicating that almost all objects in all test examples are mislabeled though only less than 80\% LL targeted mislabeling attacks succeeded.

\textbf{C. Transferability of Targeted Specificity Attacks. \/} Consider in Table~\ref{tab:online-attack-showcase} the victim detector SSD512 with the same backbone and detection algorithm as SSD300, TOG-vanishing can perfectly transfer the attack to SSD512 with the same effect (i.e., no object is detected). For TOG-fabrication, we observe that while the number of false objects is not as much as in the SSD300 case, a fairly large number of fake objects are wrongly detected by SSD512. The TOG-mislabeling (LL) attack transfers to SSD512 but with the object-fabrication effect instead, while the TOG-mislabeling (ML) attack failed to transfer for this example. Now consider YOLOv3-D and YOLOv3-M, the TOG-mislabeling (LL) attack is successful in transferability for both victims but with different attack effects, such as wrong or additional bounding boxes or wrong labels. Also, the attacks from SSD300 can successfully transfer to YOLOv3-M with different attack effects compared to the attack results in SSD300, but not to YOLOv3-D for this example. 

\end{document}